\documentclass[aps,pre,longbibliography,superscriptaddress,twocolumn,amsfonts,amsmath,amssymb,floatfix]{revtex4-1}
\usepackage[sort&compress]{natbib}

\usepackage[T1]{fontenc} % For extra glyphs (accents, etc)
\usepackage[utf8]{inputenc}

\usepackage{color}      % use if color is used in text
\usepackage{graphicx}
\usepackage{grffile}
\newlength\figurewidth
\AtBeginDocument{\setlength\figurewidth{.5\linewidth}}

\RequirePackage{siunitx}
\DeclareSIUnit\kT{\ensuremath{\mathit{k_\text{B}\!T}}}
\DeclareSIUnit\d{\ensuremath{\mathit{d}}}

\usepackage{calligra}
\usepackage{bm}
\DeclareMathAlphabet{\mathcalligra}{T1}{calligra}{m}{n}
\usepackage{bbold} % for operator identity

\allowdisplaybreaks % allows page break of aligned equations

\renewcommand{\vec}[1]{{\bm{#1}}}
\def\kT{\ensuremath{k_\text{B}T}}
\def\vFext{\ensuremath{\vec{F}_\text{ext}}}
\def\Fext{\ensuremath{F_\text{ext}}}
\def\Fc{\ensuremath{F_c}}

\newcommand{\mean}[1]{\left\langle #1 \right\rangle}

\renewcommand{\imath}{\mathrm{i}}
\newcommand{\du}{d}
\newcommand{\pderiv}[3][]{\frac{\partial^{#1}#2}{\partial {#3}^{#1}}} %total derivative

\newcommand{\vc}{\vec{c}}
\newcommand{\vh}{\vec{h}}
\newcommand{\vk}{\vec{k}}
\newcommand{\vp}{\vec{p}}
\newcommand{\vphi}{\vec{\phi}}
\newcommand{\vPhi}{\vec{\Phi}}
\newcommand{\vq}{\vec{q}}
\renewcommand{\vr}{\vec{r}}
\newcommand{\vv}{\vec{v}}
\newcommand{\vw}{\vec{w}}
\newcommand{\vx}{\vec{x}}

\newcommand{\up}{\hat{\vp}}

\renewcommand{\Re}{\operatorname{Re}}
\renewcommand{\Im}{\operatorname{Im}}

\newcommand{\re}{\text{re}}
\newcommand{\im}{\text{im}}
\newcommand{\are}{a^\re}
\newcommand{\aim}{a^\im}
\newcommand{\bre}{b^\re}
\newcommand{\bim}{b^\im} 

\newcommand{\fs}{f^s}
\newcommand{\vfs}{\vec{f}^s}

\newcommand{\cF}{\mathcal{F}}
\newcommand{\cU}{\mathcal{U}}
\newcommand{\cV}{\mathcal{V}}
\newcommand{\cW}{\mathcal{W}}
\newcommand{\cm}{\mathcalligra{m}}
\newcommand{\Landau}{\mathcal{O}}

\newcommand{\etal}{\textit{et~al. }}

\usepackage[hidelinks]{hyperref}
\def\urlprefix{}
\begin{document}
\title{Critical force in active microrheology}
\date{\today}
\def\unikn{\affiliation{%
  Fachbereich Physik, Universität Konstanz,
  78457 Konstanz, Germany}}
\def\unial{\affiliation{%
  Departamento de Física Aplicada, Universidad de Almería,
  04.120 Almería, Spain}}
\author{M.~Gruber}\unikn
\author{A.~M.~Puertas}\unial
\author{M.~Fuchs}\unikn

\begin{abstract}
Soft solids like colloidal glasses exhibit a yield stress, above which the system starts to flow.
The microscopic analogon in microrheology is the delocalization of a tracer particle
subject to an external force exceeding a threshold value, in a glassy host. We characterize this delocalization transition based on a bifurcation analysis of the corresponding mode-coupling theory equations. A schematic model is presented first, that allows analytical progress, and the full physical model is studied numerically next. This analysis yields a continuous type A transition with a critical power law decay of the probe correlation functions with exponent $-1/2$. In order to compare with simulations with a limited duration, a finite time analysis is performed, which yields reasonable results for not-too-small wave vectors. 
The theoretically predicted findings are verified by Langevin dynamics simulations. For small wave vectors we find anomalous behavior for the probe position correlation function, which can be traced back to a wave vector divergence of the critical amplitude. In addition we propose and test three methods to extract the critical force from experimental data, which provide the same value of the critical force when applied to the finite-time theory or simulations. 
\end{abstract}

\maketitle

\section{Introduction}
\label{sect_intro}

The yield stress in ductile solids is the crossover between linear elasticity and plastic
deformations. However, in soft matter systems, the yield stress is more conveniently defined as the
minimum stress needed to provoke the flow of the system \cite{Bonn2017}; in a fluid this minimum
stress is zero, while it is generically finite for a soft matter solid. Starting from simple
constitutive relations, different models have been developed describing the yield stress in terms of
the microscopic characteristics of the system \cite{Bonn2017}; in particular let us mention the soft glassy rheology model \cite{Sollich1998}, the shear
transformation zone model \cite{Falk1998} or the mode
coupling approach \cite{Fuchs2002}. Our approach to the problem is based on the application
of a localized stress, induced by a colloidal tracer that is pulled externally, namely, active microrheology.

Microrheology was proposed more than 20 years ago as a technique to access the rheological
properties at the microscopic scale, monitoring the dynamics of colloidal tracers introduced in the
sample \cite{Mason1995}. However, it was soon acknowledged that the technique could be significantly
improved if the tracer is pulled externally (active microrheology), as both the linear and
non-linear regimes can be studied \cite{Puertas2014,Furst2017}. Several models for active microrheology have been presented, based on an effective medium approach \cite{Levine2000}, the two-particle Smoluchowski equation (for low densities) for stationary \cite{Squires2005} and transient regimes \cite{Leitmann2018}, the mode-coupling approach (applicable to high densities) \cite{Gazuz2009}, the continuous time random walk model \cite{Schroer2013,Gradenigo2016}, a kinetically constrained model \cite{Jack2008,Gradenigo2016}. Simulation studies \cite{Williams2006,Winter2012,Reichhardt2015} and  first experimental studies confirmed that the dynamics becomes highly anomalous in glass-forming dispersions \cite{Habdas2004,Senbil2019}.

In a previous study, active microrheology in a colloidal glass was analyzed with a model based on the
mode-coupling approximation, and tested against simulations \cite{Gruber2016}. It was found in that
work that there are two regimes for the dynamics of the tracer: a localized regime, found when the
external force is small, and where the tracer is trapped in the cage formed by its own neighbours,
and a delocalized regime, when the force is large enough to break the cage, and the tracer exhibits
motion over long distances. The properties of the localized regime were studied in detail in that
work, and confirmed by the simulations. The focus of the present study, is the examination of the system's behavior at the crossover between these two regimes.

We present in this paper the properties of the critical force separating both regimes within the model, and use them
to nail down the critical force in the simulations and confirm the predictions of the model. In
particular, it is predicted that the long-time limit of the tracer position correlation function (nonergodicity parameter)
decays linearly with the external force, being zero at the critical force (identified as a
\emph{type A} transition within mode-coupling theory). At the critical force, the correlation
function decays as function of time according to a power-law with exponent $-1/2$, and a prefactor that coincides with the slope of the
nonergodicity parameter with the external force.  For small wave vectors perpendicular to the
external force some anomalies are found in the theory. 
Because the long time limit is generally unreachable in the simulations, we test within the theory, how the predictions are altered if values
available at finite times are considered. Our simulation results confirm the predictions of the
model, validating the analysis and providing a toolbox to estimate the critical force. This can prove useful to identify the critical force in experimental systems \cite{Senbil2019}. 

\section{Theory}
\label{theory}

In active microrheology we consider a spherical \emph{probe} particle with diameter $d$, which is pulled
by a constant external force $\vFext = \Fext \hat{\vec{e}}_z$ through a colloidal
suspension of spheres of the same diameter, the \emph{host}. All particles are subject to Brownian motion induced by the suspending
fluid, but hydrodynamic interactions are neglected. Our description is based on the
displacement distribution function $G^s(\vr,t)$ (also called the self-part of the
van-Hove-function) and, more precise, on its spatial Fourier transform
\begin{equation}
 \Phi^s_\vq(t) = \int e^{i \vq\cdot\vr} G^s(\vr,t)\du \vr,
\end{equation}
also called probe correlator. 

\subsection{Model}
The dynamics of this probe correlation function is derived from a microscopic overdamped
Smoluchowski equation within the framework of mode-coupling theory (MCT) \cite{Gruber2016}. It
is given by a integrodifferential equation
\begin{equation}
 0 = \partial_t \Phi^s_\vq(t) + \Gamma_\vq \Phi^s_\vq(t) + \int_0^t
 \cm_\vq(t-t')\partial_{t'}\Phi^s_\vq(t')\du t'
 \label{eq:eom_probe_correlator}
\end{equation}
with initial condition $\Phi^s_\vq(0)=1$. $\Gamma_\vq=q^2-\imath \vq \Fext$ describes the free decay of the correlator as
if no other particles were present and the memory kernel $\cm_\vq$ accounts for the interactions
between the probe and the bath. $\cm_\vq$ is given by a functional of the probe
and the host correlators using MCT. 
For the scaling analysis it is convenient to introduce the Laplace transformed quantities given by
\begin{equation}
 \tilde{A}(s) = \int_0^\infty \du t e^{-st}A(t).
\end{equation}
The equation of motion \eqref{eq:eom_probe_correlator} then reads
% \begin{equation}
% 0 = (s\tilde{\Phi}(s)-1) + \Gamma_\vq \tilde{\Phi}(s) + \tilde{\cm}_\vq(s) (s\tilde{\Phi}(s)-1)
% \end{equation}
% or 
\begin{equation}
\tilde{\Phi}^s_\vq(s) = \left(s + \frac{\Gamma_\vq}{1+\tilde{\cm}_\vq(s)}\right)^{-1}.
\label{eq:eom_probe_correlator_laplace}
\end{equation}
Using the cylindrical symmetry of the system around the force direction, we find (choosing $\vq = (q_x,0,q_z)$ without loss of
generality)
\begin{equation}
\label{eq:effective_memory}
 \frac{\Gamma_\vq}{1+\tilde{\cm}_\vq} = \frac{\Gamma_\vq^x(1+\tilde{m}_\vq^{zz}) -
 \Gamma_\vq^{xz}\tilde{m}_\vq^{xz} +
 \Gamma_\vq^z(1+\tilde{m}_\vq^{xx})}{(1+\tilde{m}_\vq^{xx})(1+\tilde{m}_\vq^{zz}) -
 \tilde{m}_\vq^{xz}\tilde{m}_\vq^{xz}}
\end{equation}
with $\Gamma_\vq^x = q_x^2, \Gamma_\vq^z = q_z^2 - i q_z \Fext$, $\Gamma_\vq^{xz} = 2q_xq_z-i q_x
\Fext$ and the primitive memory functionals
\begin{equation}
 m_\vq^{\alpha\beta}[\vPhi^s(t),\vPhi(t)] = \frac{1}{(2\pi)^3} \int \du \vk p_\alpha p_\beta
 \frac{(S^s_{p})^2}{n S_{p}}\Phi^s_\vk(t)\Phi_{p}(t)
 \label{eq:primitive_memory_functional_time}
\end{equation}
with $\vp=\vq-\vk$. $S_p$ is the static bath structure factor, $S^s_p$ the static probe-bath structure factor and $\Phi_p(t)$ the
bath correlation function. With the structure factor obtained in the Percus-Yevick approximation for hard spheres, the host presents a glass transition within MCT at $\varphi_g=\num{.516}$. For this work, we chose a packing fraction of $\varphi=\num{.537}$, which is about \SI{4}{\percent} above the glass transition.

\subsection{Numerical details}
The numerical solution algorithm is described in Appendix C of \cite{Gruber2016} and in more detail
in Chapter 3 of \cite{GruberPhD2019}. We choose a cutoff of $q_{\max}d = 14$ for $q_r$ and $q_z$ on
a uniform grid from 0 to $q_{\max}d$ with step size $\Delta q d = 0.5$ with $N=29$ points, which is
refined towards 0 by adding $N_\text{log}=10$ nonuniformly spaced grid points given by $2^{-i} \Delta q d$
($i=1,\ldots,10$). This choice allows sufficient resolution for long-ranged structures as well as
the microscopic structures, while allowing reasonable computing times ($39\times 39$
correlation functions have to be calculated) and preventing numerical issues \footnote{For a cutoff
$q_\text{max}d \geq 20$ the transition type changes, see Chapter 5 in \cite{GruberPhD2019}}. The
dynamical solutions are obtained using the decimation algorithm with an initial step size of
$\Delta t d^2/D_0 = 10^{-8}$ on a grid with 1024 grid points using up to 45 decimation steps. The
bath correlator is calculated on the same time grid and the usual grid in $q$-space with cutoff 
$qd = 65$ and 512 grid points. For these parameters (and a packing fraction of $\varphi=\num{.537}$)
the critical force is given by $F_c= \SI{44.79+-0.01}{\kT/\d}$. For the analysis of the critical
dynamics, we use $N_\text{log}=25$ points in the nonuniform part of the grid for a better resolution
of the critical force, which is given by $F_c = \SI{44.7815+-0.001}{\kT/\d}$. We tested two algorithms for the time integration based on (i) the integral equation and (ii) the integro-differential equation representation for the effective memory function. They give the same results, but differ in the regions of stability, see Chapter~3 of \cite{GruberPhD2019} for details.

\section{Bifurcation analysis}
\label{bifurcation}
In this section, we investigate the behavior of the probe correlation function $\Phi^s_\vq(t)$ close
to the delocalization transition, viz.~a kind of depinning transition. This transition is characterized by the
long-time limits $f^s_\vq:= \lim_{t\to\infty}\Phi^s_\vq(t)$. In the glass, i.e.\ for $\varphi>\varphi_g$, we find
$f^s_\vq \neq 0$ for small external forces, while $f^s_\vq = 0$ if the force is large enough. The
smallest force for which $f^s_\vq = 0$ is the critical force and determines the delocalization
transition \cite{Gruber2016}. The predictions for the behavior of the correlation function close to
this critical point will provide us with several means to characterize the critical force in
simulations and experiments.

Starting point for the bifurcation analysis is the Laplace space version of the equations of motion
\eqref{eq:eom_probe_correlator_laplace}. In order to make analytic progress, we approximate the bath
correlation function by its long time limit $\Phi_q(t) \approx f_q:= \lim_{t\to\infty} \Phi_q(t)$.
Then, the primitive memory functionals become linear in $\vPhi^s(t)$ so that we can directly
insert the Laplace transformed tagged-particle correlator $\tilde{\Phi}^s_q(s)$. To facilitate a
discussion of the full equations, we will first perform the bifurcation analysis of a
simplified version, a schematic model. 

\subsection{Schematic model}
A schematic model corresponding to the full MCT equations was designed by Gustavo
Abade modifying previous schematic models \cite{Gnann2011,Gazuz2013}. It focuses on the characteristic time-dependent behavior of the (complex-valued) parallel and (real-valued) perpendicular
modes of the probe-bath correlation function. In the following they are summarized in the vector $\vphi(t)=(\phi_\parallel(t),\phi_\perp(t))$ with $\phi_\parallel(t)\in\mathbb{C}$ and $\phi_\perp(t)\in\mathbb{R}$. 
The equations of motion are given by
\begin{equation}
  \tau_i \partial_t \phi_i + \phi_i(t) + \int_0^t
  m_i(t-t')\partial_{t'}\phi_i(t')\du{t'} = 0
  \label{eq:schematic:eom}
\end{equation}
for $i\in\{\parallel,\perp\}$. They have the same structure as the full model in \eqref{eq:eom_probe_correlator} (multiplied by the timescale $\tau_i=\Gamma_i^{-1}$). This includes the complex valued relaxation times for the tagged-particle correlator $\tau_{\parallel} = \tau^s\left(1 - \imath \kappa_{\parallel} \Fext\right)^{-1}$ and $\tau_{\perp} = \tau^s$, where $\tau^s$ describes the relaxation time without external force. 

To mimic the couplings introduced by the memory functionals, we have to find an appropriate simplification of \eqref{eq:effective_memory}. For the parallel direction (i.e.\ $q_x=0$) it reduces to the condition $\cm_{\vq} = m^{zz}_{\vq}$ so that we can define for the parallel direction $m_\parallel(t) = \cF_\parallel(\vphi(t))$ as in Ref.~\cite{Gnann2011} via
\begin{align}
 \cF_\parallel(\vx) &=  \frac{(v_1^s x_\parallel^\ast + v_2^s x_\perp) f_b}{1
 - \imath\kappa_\parallel \Fext},
 \end{align}
where $v_1^s$ and $v_2^s$ describe the coupling of the perpendicular and the
parallel component to the bath mode, while $\kappa_\perp$ (appearing in the next
equation) and $\kappa_\parallel$ describe the coupling of the probe correlators
to the force. $f_b$ is the nonergodicity parameter of the $F_{12}$ model and describes the frozen-in glass structure of the bath.
The star $^\ast$ denotes complex conjugation. Hereby, we just replaced the memory functional by a simple polynomial, which is the standard procedure for schematic MCT models (see e.g.\ p.~202ff in \cite{Goetze2009}). Furthermore, this memory functional coincides with the parallel memory functional of the schematic model discussed by Gnann \etal in Ref.~\cite{Gnann2011}, which is itself an extension of the one-component schematic model of Gazuz  \cite{Gazuz2013}.

For the perpendicular direction (i.e.\ $q_z=0$), the equation of motion \eqref{eq:effective_memory} for the effective memory function $m_\perp(t)$ does not reduce to a simple memory function as was the case in the model in Ref.~\cite{Gnann2011}. Instead, we obtain a nonlocal integro-differential equation of the form 
 \begin{align}
 &\,\tau_\perp m_\perp(t) + \int_0^t m_\perp(t-t') \left(m_\perp^{zz}(t) + \imath
 \kappa_\perp\Fext m_\perp^{xz}(t)\right)\notag\\ 
 = &\, 
 \tau_\perp\left(m_\perp^{xx}(t) - \imath \kappa_\perp\Fext m_\perp^{xz}(t) \right)
 + \int_0^t m_\perp^{xx}(t-t')m_\perp^{zz}(t')\du{t'} \notag\\
 &\, -  \int_0^t m_\perp^{xz}(t-t')  m_\perp^{xz}(t')\du{t'},
\label{eq:schematic:nonlocal_memory_kernel}
\end{align}
with the local memory functionals $m_\perp^{\alpha\beta}(t) = \cF_\perp^{\alpha\beta}(\vphi(t))$ given by
\begin{subequations}
\label{eq:schematic:memory_perp}
\begin{align}
 \cF_\perp^{xx}(\vx) &= (v_1^s x_\perp + v_2^s \Re\{x_\parallel\}) f_b,\\
 \cF_\perp^{xz}(\vx) &= -\imath v_2^s \Im\{x_\parallel\} f_b, \\ 
 \cF_\perp^{zz}(\vx) &= v_2^s \Re\{x_\parallel\} f_b.
\end{align}
\end{subequations}
Since the perpendicular correlation function is real valued, this requires the effective perpendicular memory function to be real as well. The symmetries of the full primitive memory functionals suggest the given couplings to the real or imaginary part only. 

The schematic model defined by \eqref{eq:schematic:eom}-\eqref{eq:schematic:memory_perp} has the free parameters $\tau^s$, $\kappa_\parallel$, $\kappa_\perp$, $\Fext$, $v_1^s$, $v_2^s$ and $f_b$. The bath nonergodicity parameter $f_b$  will be parametrized through the distance $\varepsilon$ to the critical point (viz.~glass transition) \footnote{We choose as parameters for the $F_{12}$-model $v_2=2$ and $v_1=2(\sqrt{2}-1) + \varepsilon (\sqrt{2}-1)^{-1}$. This choice of the parameters yields for $\varepsilon>0$ the bath nonergodicity parameter 
$f_b(\varepsilon) = \bigl(4-2\sqrt{2}-\varepsilon/(\sqrt{2}-1) + \sqrt{\varepsilon^2(3+2\sqrt{2})+4\varepsilon(2+\sqrt{2}}\bigr)/4$, where $\varepsilon$ determines the distance from the glass transition \cite{Gazuz2013}}.
%$f_b=(v_2-v_1 + \sqrt{(v_2-v_1)^2-4(1-v_1)v_2})/(2v_2)$ \noteMF{give relation here} 
% given by $v_2=2$ and $v_1=2(\sqrt{2}-1) + \varepsilon (\sqrt{2}-1)^{-1}$ 
The parameter $\Fext$ will be used as analogon to the external force in the full model to drive the system through the delocalization transition. Time will be expressed in multiples of $\tau^s$, which is equivalent to setting $\tau^s = 1$.

\subsection{Beta-scaling analysis of the schematic model}
For the beta-scaling analysis, we rewrite the equations of motion in terms of the real
valued quantities $\phi_1:=\Re \phi_\parallel$, $\phi_2:=\Im \phi_\parallel$ and
$\phi_3:=\phi_\perp$ and obtain in Laplace space
\begin{subequations}
\label{eq:schematic:full_eq_motion}
\begin{align}
- \frac{\phi_{1}(s) \left(\phi_{1}(s) s - 1\right) + \phi_{2}^{2}(s) s}{\phi_{2}^{2}(s) s^{2} +
\left(\phi_{1}(s) s - 1\right)^{2}} 
= \sum_{i=1}^3 \left(V_{{\parallel}}^{\mathrm{re}}\right)_i \phi_i(s) + {\tau_1},
\label{eq:schematic:full_eq_motion_1}\\
\frac{\phi_{2}(s)}{\phi_{2}^{2}(s) s^{2} + \left(\phi_{1}(s) s - 1\right)^{2}}
= \sum_{i=1}^3 \left(V_{\parallel}^{\mathrm{im}}\right)_i \phi_i(s) + {\tau_2},
\label{eq:schematic:full_eq_motion_2} \\
\frac{\phi_3(s)}{1 - s\phi_3(s)} = 
\frac{\sum_{i,j=1}^3 \left(W_{{\perp}}\right)_{ij} \phi_i(s) \phi_j(s)}{\tau_3 + \sum_{i=1}^3
\left(U_{\perp}\right)_i \phi_i(s)}\notag\\
 + \frac{\sum_{i=1}^3 \left(V_{\perp
}\right)_i \phi_i(s)}{\tau_3 + \sum_{i=1}^3 \left(U_{\perp}\right)_i \phi_i(s)}+ \tau_3.
\label{eq:schematic:full_eq_motion_3}
\end{align}
\end{subequations}
Note that the tildes to indicate Laplace space values have
been dropped for the ease of notation. The scalars $\tau_i$, the vectors $V^\text{re}_\parallel$,
$V^\text{im}_\parallel$, $V_\perp$, $U_\perp$ and the symmetric matrix $W_\perp$ contain
combinations of the model parameters as introduced in Chapter 6 of \cite{GruberPhD2019} and summarized in Appendix A. 

Using the Laplace limit theorem $\lim_{t\to\infty} \phi_i(t) = \lim_{s\to 0} s \tilde{\phi}_i(s)$ we
can identify the determining equation for the nonergodicity parameters $f_i:=\lim_{t\to\infty}\phi_i(t)$
by multiplication of \eqref{eq:schematic:full_eq_motion} by $s$ and taking the limit $s\to 0$. The
nonergodicity parameters are then given by the roots of the following set of nonlinear equations
\begin{subequations}
\label{eq:schematic:nonergodicity_root_equation}
\begin{align}
 J_1(\vx) &= - \frac{x_1 \left(x_1 - 1\right) + x_2^2}{x_2^2 +
\left(x_1 - 1\right)^{2}} - \sum_{i=1}^3
\left(V_{{\parallel}}^{\mathrm{re}}\right)_i x_i,\\
 J_2(\vx) &= \frac{x_2}{x_2^2 + \left(x_1 - 1\right)^{2}} - 
\sum_{i=1}^3 \left(V_{\parallel}^{\mathrm{im}}\right)_i x_i,\\ 
 J_3(\vx) &= \frac{x_3}{1 - x_3} -
\frac{\sum_{i,j=1}^3 \left(W_{{\perp}}\right)_{ij} x_i x_j }{\sum_{i=1}^3 \left(U_{\perp}\right)_i
x_i}.
\end{align}
\end{subequations}

\begin{figure}
%TODO: adjust font size
\includegraphics[width=0.95\figurewidth]{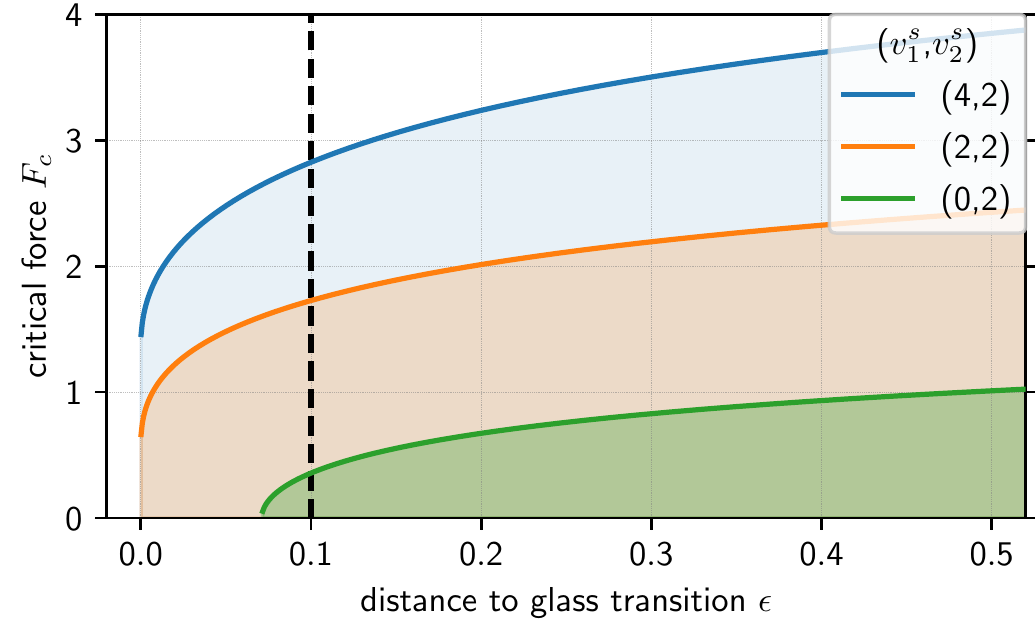}
\caption{Phase diagram of the schematic model with $\kappa_\parallel=\kappa_\perp=1$ for different couplings. In the shaded regions below the solid lines, the solutions are nonergodic, while they are ergodic above. Different values of $v_1^s$ for $v_2^s=2$ are presented, as labeled. According to \eqref{eq:schematic:critical_force} there are only quantitative changes on varying $v_2^s$, but the shape is not affected.
 \label{schematic_phase_diagram}}
\end{figure}
In the glass ($\varepsilon>0$) for vanishing force $\Fext$ and for not too small couplings $v^s_1$, $v^s_2$, there is a nontrivial solution $f_i\neq 0$ of this set of eqns.~\eqref{eq:schematic:nonergodicity_root_equation}. Increasing the force, there is a critical force $\Fc$, above which there exists only the trivial solution $f_i=0$. This is visualized for some parameters in Fig.~\ref{schematic_phase_diagram}. If the coupling to the bath is too weak (e.g.\ $v^s_1=0, v^s_2=2$),  the critical force appears only deep in the glass. The lines for the critical forces for $\kappa_\parallel=\kappa_\perp=\kappa$ were obtained analytically in Section~6.2.4 in \cite{GruberPhD2019}, which is reproduced in Appendix~\ref{sec:critical_force} and read 
\begin{equation}
 \Fc = \frac{1}{\kappa}\left(\frac{\left(f_b v_2^s\right)^2}{2}\left(2 \beta^2 +
 1 + \sqrt{8\beta^2 + 1 + \frac{8\beta}{f_b v_2^s}}\right) - 1\right)^{\frac{1}{2}}
 \label{eq:schematic:critical_force}
\end{equation}
with $\beta = v_1^s/v_2^s$. 

Critical points are determined by the set of parameters at which two (or more) roots coalesce. This
implies that the Jacobian of \eqref{eq:schematic:nonergodicity_root_equation} is not invertible.
Anticipating that the nonergodicity parameters will vanish at the critical point, we find for the
stability matrix $S^c(\vx)_{ij} = \partial_{x_j}J_i^c(\vx)$ (Jacobian matrix at the critical point)
\begin{subequations}
\label{eq:schematic:critical_stability_matrix}
\begin{align}
 S^c(\vx)_{1j} &= \delta_{1j} 
 	- \left(V_{{\parallel}}^{\mathrm{re}}\right)_j, \\
 S^c(\vx)_{2j} &= \delta_{2j}
    - \left(V_{\parallel}^{\mathrm{im}}\right)_j,
 \\
 S^c(\vx)_{3j} &= \delta_{3j}
  - \frac{2\sum_{i=1}^3\left(W_{{\perp}}\right)_{ij} x_i - x_3
  \left(U_{\perp}\right)_j}{\sum_{i=1}^3 \left(U_{\perp}\right)_i x_i}.
\end{align}
\end{subequations} 
This representation makes use of $x_i\ll 1$, the symmetry of $W_\perp$ and the condition for the root $F_3^c(\vx)=0$. In classical MCT, the stability matrix is only a function of the coupling coefficients. This allowed it to reformulate the condition of a non-invertible Jacobian into the problem of finding an eigenvector of the stability matrix corresponding to the eigenvalue 0. In our case here, however, the stability matrix depends additionally on the critical nonergodicity parameter $\vx$. As a consequence, the linear problem of finding an eigenvector to the eigenvalue 0 of the stability matrix transforms into solving the nonlinear problem $S^c(\vh)\vh = 0$. This avoids the problem of determining the limit $|\vx|\to 0$ of the stability matrix $S^c(\vx)$, because the representation is scale-free, i.e.\ any scalar multiple of $\vh$ is a solution as well. 
See Appendix B where the critical force (Eq.~\ref{eq:schematic:critical_force}) is derived following this strategy.

\begin{figure}
\includegraphics[width=0.95\figurewidth]{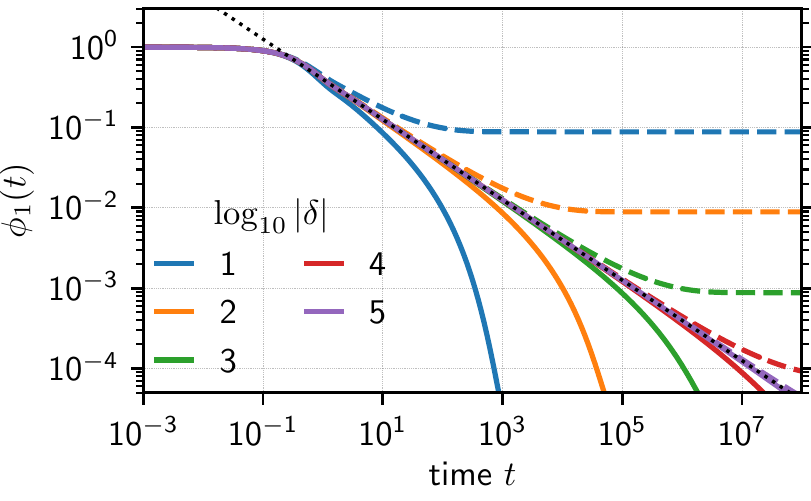}
\includegraphics[width=0.95\figurewidth]{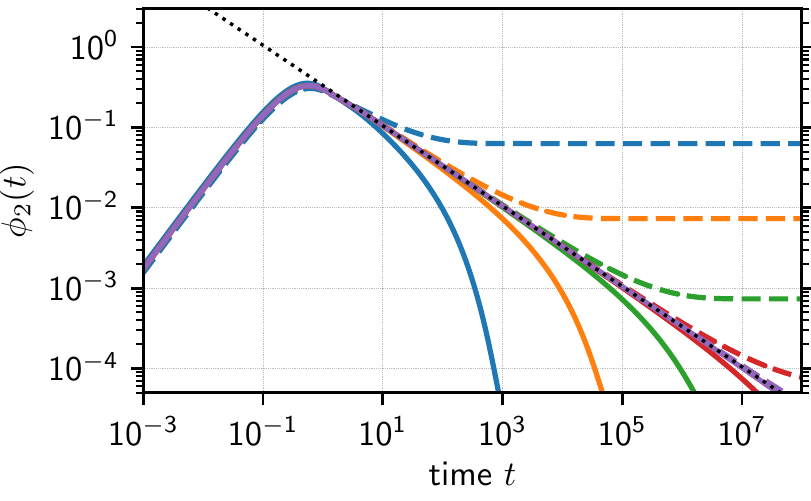}
\includegraphics[width=0.95\figurewidth]{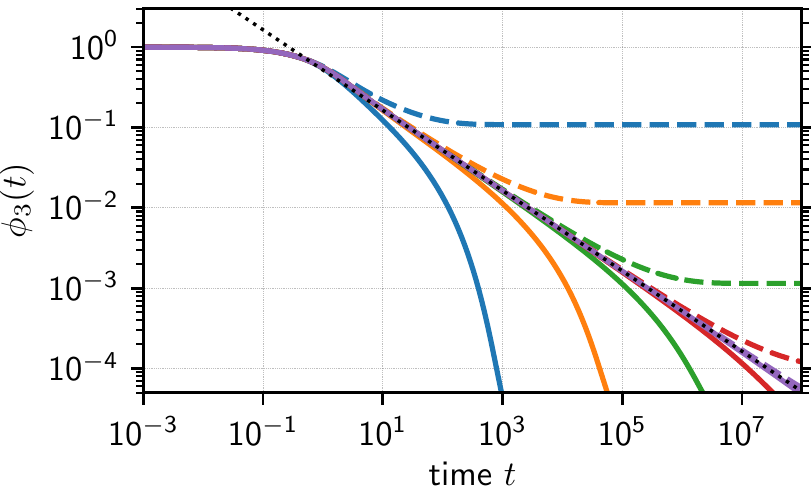}
\caption{Critical behavior of the schematic model. Correlation functions for $\epsilon=\num{.1}$, $v_1^s=v_2^s=\num{2}$,
  $\kappa_\parallel=\kappa_\perp = 1$, constant bath and different distances
  $\delta=(\Fext-\Fc)/\Fc$ to the critical force as labeled.
  Dashed lines indicate $\delta<0$, while solid lines indicate $\delta>0$. The
  critical law $\phi_i^c(t) = \alpha h_i t^{-1/2}$ is shown as dotted line. The critical amplitude $h_i$ is a solution of \eqref{eq:schematic:fixed_point_eq_hcrit} and its magnitude $\alpha$ is adjusted based on the data in the top panel. The line color and style is the same across all panels.
 \label{critical_law_schematic_correlators}}
\end{figure}

For the analysis of the critical (long-time) dynamics, we expand
\eqref{eq:schematic:full_eq_motion} for small $s$ and large $\phi_i(s)$
%\noteAP{Is this correct? I am curious because the correlators decay and at long times, small $s$, are close to zero. } \noteMG{I added the argument $s$ to clarify the following: If $\phi_i(t)$ approaches a plateau (even a small one) for large times, we obtain a $1/s$ pole for $\phi_i(s)$ for $s\to 0$. There holds $\phi_i(s\to 0)=\int_0^\infty \phi_i(t)\du t$. This implies large values of $\phi_i(s)$ for slowly decaying correlation functions - even if they decay to 0 finally.} 
and find
% \begin{widetext}
\begin{subequations}
\label{eq:eom_schematic_critical}
\begin{align}
\phi_1(s)  + s(\phi_1^2(s) - \phi_2^2(s)) &\approx  \sum_{i=1}^3
\left(V_{\parallel}^{\mathrm{re}}\right)_i \phi_i(s) + {\tau_1},\\
\phi_2(s) + 2 s \phi_1(s) \phi_2(s) &\approx 
\sum_{i=1}^3 \left(V_{\parallel}^{\mathrm{im}}\right)_i \phi_i(s) + \tau_2,\\
\phi_3(s) + s\phi_3^2(s) &\approx \frac{\sum_{i,j=1}^3(W_\perp)_{ij}\phi_i(s)\phi_j(s))}{\sum_{i=1}^3(U_\perp)_i
\phi_i(s)} \notag\\
& + \frac{\sum_{i=1}^3(V_\perp)_i\phi_i(s)}{\sum_{i=1}^3(U_\perp)_i
\phi_i(s)} + \tau_3 \notag\\
& - \tau_3\frac{\sum_{i,j=1}^3(W_\perp)_{ij}\phi_i(s)\phi_j(s))}{\left(\sum_{i=1}^3(U_\perp)_i
\phi_i(s)\right)^2}.
\end{align}
\end{subequations}
% \end{widetext}
We can now use the ansatz for an approximate solution of $\phi_i(s)$
\begin{equation}
\label{eq:schematic:ansatz}
 \phi_i(s) = h_i s^{-1/2} + c_i + \Landau(s^{1/2})
\end{equation}
to derive equations for the unknown parameters $h_i$, the so called \emph{critical amplitudes}.
This ansatz implies $s^2\phi^3(s)=\Landau(s^{1/2})$ and $\phi^{-1}(s) = \Landau(s^{1/2})$ and yields
in first order
\begin{subequations}
\begin{align}
0 &= s^{-1/2}\left(h_1 - \sum_{i=1}^3
\left(V_{\parallel}^{\mathrm{re}}\right)_i h_i\right) 
 %+ \left(c_1 + h_1^2 - h_2^2 - \sum_{i=1}^3 \left(V_{\parallel}^{\mathrm{re}}\right)_i c_i -
%\tau_1 \right) + \Landau(s^{1/2}),
+ \Landau(s^0),
\\
0 &= s^{-1/2}\left(h_2 - \sum_{i=1}^3
\left(V_{\parallel}^{\mathrm{im}}\right)_i h_i\right) 
%+ \left(c_2 + 2 h_1 h_2 - \sum_{i=1}^3 \left(V_{\parallel}^{\mathrm{im}}\right)_i c_i - \tau_2
% \right) + \Landau(s^{1/2}), 
+ \Landau(s^0),
\\
0 &= s^{-1/2}\left(h_3 - \frac{\sum_{i,j=1}^3 \left(W_{\perp}\right)_{ij} h_i
h_j}{\sum_{i=1}^3 \left(U_{\perp}\right)_i h_i}\right)
+ \Landau(s^0). 
% \\ &\quad + \left(c_3 + h_3^2 - \frac{\sum_{i,j\in
% I}2\left(W_{\perp}\right)_{ij} h_i c_j + \sum_{i=1}^3 \left(V_{\perp}\right)_ih_i -
% h_3(\tau_3 + \sum_i \left(U_{\perp}\right)_i c_i)}{\sum_{i=1}^3 \left(U_{\perp}\right)_i h_i}
% - \tau_3 \right) + \Landau(s^{1/2}), \notag
\end{align}
\end{subequations}
Solving this set of equations for arbitrary $s$ up to order $\Landau(s^{-1/2})$ requires $\vh$, the
vector of critical amplitudes, to fulfill the following fixed-point equation
\begin{subequations}
\label{eq:schematic:fixed_point_eq_hcrit}
\begin{align}
h_1 &= \sum_{i=1}^3 \left(V_{\parallel}^{\mathrm{re}}\right)_i h_i,\\ 
h_2 &= \sum_{i=1}^3 \left(V_{\parallel}^{\mathrm{im}}\right)_i h_i,\\ 
h_3 &= \frac{\sum_{i,j=1}^3 \left(W_{\perp}\right)_{ij} h_i h_j}{\sum_{i=1}^3
\left(U_{\perp}\right)_i h_i}.
\label{eq:schematic:fixed_point_eq_h3}
\end{align}
\end{subequations}
This equation is scale free with respect to $\vh$, since any scalar multiple of $\vh$ is again a
solution. The magnitude of $\vh$ can be fixed using the higher order terms to determine the vector
$\vc$ introduced in the expansion Eq.~\eqref{eq:schematic:ansatz} as worked out in Chapter~6 of \cite{GruberPhD2019}. The critical exponent of $-1/2$ in Laplace space translates into a critical power law in time space with exponent $-1/2$ as well. This critical behavior can be verified numerically as shown in Fig.~\ref{critical_law_schematic_correlators}. Close to the critical point this power law can be observed over more than six decades in the numerical solutions.

As anticipated above, $\vh$ is an
eigenvector of the critical stability matrix $S^c(\vh)$ to the eigenvalue 0, i.e.\ $S^c(\vh)\vh = 0$, which can be verified by direct calculation. 
%This fixed-point equation can be used to find the analytical expression \eqref{eq:schematic:critical_force} for the critical force in the case $\kappa_\parallel = \kappa_\perp$ as shown
%in Chapter~6.2.4 of \cite{GruberPhD2019}. The corresponding phase diagram is shown in Fig.~\ref{schematic_phase_diagram} 
A Taylor expansion of this solution is possible and leads to a continuous type A transition, which can be confirmed numerically. Inverting \eqref{eq:schematic:ansatz} to find the critical behavior in the time domain, we
corroborate the resulting $s^{-1/2}$ power law numerically, including the prefactors determined by
$\vh$. This is shown in Fig.~\ref{critical_law_schematic_correlators}, where we plot the numerical solutions for decreasing but finite distances $|\delta|$ from the critical point. The critical law $\phi_i(t)=\alpha h_i t^{-1/2}$ is shown as black dotted line, where the magnitude $\alpha$ has been adjusted to the data for $i=1$. 

In Appendix~\ref{sec:scaling_memory_kernel} we derive a scaling law for the time evolution close to the critical point. It connects the timescale for the deviation from the critical law to the inverse square of the relative distance from the critical force, i.e.\ $\tau = \delta^{-2}$. This can be verified in the numerical solutions shown in Fig.~\ref{critical_law_schematic_correlators}.

To conclude, the beta-scaling analysis of this schematic model results in a continuous transition with a
critical power law with exponent $-1/2$ in time and Laplace domain. A special point is that the eigenvector equation $S^c(\vh)\vh = 0$ for the stability matrix becomes a nonlinear problem. This leads to a scale free fixed-point equation for the critical amplitude $\vh$.

\subsection{Beta-scaling analysis of the full model}
For the analysis of the full model, we will interpret the wave vector as a discrete variable (i.e.\
$\Phi^s(t)\in\mathbb{C}^n$) instead of a continuous one ($\Phi^s(t): \mathbb{R}^3\to\mathbb{C}$).
This is also done for the numerical solution and simplifies some of the arguments. The discrete 
wave vectors will be labeled by the index $q$. As a consequence, the equations of motion become a
finite dimensional system of equations and the primitive memory functionals can be expressed as
matrices. As before, we use the constant-bath approximation $\Phi_q(t) \to f_q$ so that we can
rewrite the equations of motion as follows (the argument $s$ is skipped to keep the equation
concise)
\begin{equation}
\label{eq:eom_laplace}
\begin{split}
 0 =& (s\tilde{\Phi}_q^s - 1) + \frac{1}{s}\left(s\tilde{\Phi}_q^s
 \Gamma_q + (s\tilde{\Phi}_q^s-1)C_q^T s\tilde{\vPhi}^s\right) \\
 &+ \frac{1}{s^2}\left((s\tilde{\Phi}_q^s - 1) (s\tilde{\vPhi}^s)^T A_q
 (s\tilde{\vPhi}^s) + s \tilde{\Phi}_q^s B_q^T (s\tilde{\vPhi}^s) \right)
 \end{split}
\end{equation}
with the abbreviations
\begin{subequations}
\begin{align}
 \vv^T A_q \vw =& \frac{1}{2}\left(\mu_q^{xx}(\vv)\mu_q^{zz}(\vw) +
 \mu_q^{xx}(\vw)\mu_q^{zz}(\vv)\right) \nonumber\\
 &- (\mu_q^{xz}(\vv)\mu_q^{xz}(\vw)),\\
 B_q^T \vv =& \mu_q^{xx}(\vv)q_z^2 + \mu_q^{xz}(\vv)(-2q_x q_z) + \mu_q^{zz}(\vv)q_x^2
 \nonumber\\
 & + \imath\Fext \left(-\mu_q^{xx}(\vv)q_z + \mu_q^{xz}(\vv)q_x\right),\\
 C_q^T\vv =& \mu_q^{xx}(\vv) + \mu_q^{zz}(\vv),\\
 \Gamma_q =& q_x^2 + q_z^2 - \imath\Fext q_z
\end{align} 
\end{subequations} 
using matrix-vector multiplications with auxiliary vectors $\vv$ and $\vw$. From \eqref{eq:primitive_memory_functional_time} we obtain the long-time-limits of the memory functionals via $\mu^{\alpha\beta}_q(\vv):=m_q^{\alpha\beta}[\vv,\vec{f}]$ with the bath nonergodicity parameter $\vec{f}:=\lim_{t\to\infty} \vphi(t)$. $A_q$ is a matrix with constant coefficients, $B_q$ and
$C_q$ are vectors with constant coefficients and $\Gamma_q$ is a scalar for each wave vector,
which is labeled by the index $q$. Notice that only $B_q^T$ and $\Gamma_q$ depend on the external
force.
In order to find the proper long time behavior it is convenient to introduce the $S$-transform as
$s$ times the Laplace-transform ($s$ being the variable in Laplace space).
This has the advantage that the power-laws in the time-domain and for the $s$-transform are simply
reciprocals of each other and constant functions are transformed into constant functions. This
rationalizes the notation $s^{-1}(s\tilde{\Phi}^s_q)$ instead of writing $\tilde{\Phi}^s_q$ only.

As above, we can find the determining
equation for the nonergodicity parameters $\fs_q := \lim_{t\to\infty} \phi^s_q(t) = \lim_{s\to 0} s \tilde{\Phi}^s_q(s)$
by multiplying with $s^2$ and taking the limit $s\to 0$. It reads 
\begin{equation}
0 = (\fs_q - 1)((\vfs)^T A_q \vfs) + \fs_q (B_q^T \vfs).
\label{eq:nonergodicity_equation}
\end{equation}
The corresponding Jacobian is given by 
\begin{equation}
\begin{split}
    S_{qp} =& \delta_{qp}((\vfs)^T A_q \vfs + B_q^T \vfs) + \fs_q B_q^T
        \up
        \\
        &+ (\fs_q - 1) \left( (\vfs)^T A_q \up + (\up)^T A_q \vfs \right),
\end{split}
\end{equation}
where $\up$ is the unit vector in $p$-direction.
This matrix is zero in the limit $\vfs\to 0$. 

Equation~\eqref{eq:nonergodicity_equation} can be solved for $\fs_q/(1-\fs_q)$ to have the same form as in classical MCT. Then, the right hand side corresponds to the memory functional, which is always a polynomial in classical MCT. In our case, the resulting right hand is a quotient of the memory functionals $(\vfs)^T A_q \vfs$ and $B_q^T \vfs$. This poses a challenge for taking the limit $|\fs_q|\to 0$, as the limiting value of the quotient will depend on the details of how the different values of $\fs_q$ approach zero.

It is also possible to rewrite \eqref{eq:nonergodicity_equation} in an alternative form as a fixed-point problem 
\begin{equation}
 0 = \fs_q - \frac{(\vfs)^T A_q \vfs}{B_q^T \vfs + (\vfs)^T A_q \vfs}.
 \label{eq:nonergodicity_equation_fp} 
\end{equation}
%This implies the use of the following Jacobian
%\begin{equation}
%\label{eq:beta:stability_matrix_quotient}
% S'_{qp} = \delta_{qp} - (1-f^s_q)^2 \left(\frac{\up^T A_q\vfs +
% (\vfs)^T A_q \up}{B_q^T\vfs} - \frac{\vfs A_q \vfs}{(B_q^T
% \vfs)^2} B_q^T\up\right).
%\end{equation}
Since $\vfs$ is complex valued, the calculation of the Jacobian becomes a rather tedious task, as real and imaginary parts have to be treated as separate variables (see Appendix~\ref{chap:stability_matrix_full_model} or equations (5.37)-(5.41) in Chapter~5.4 of \cite{GruberPhD2019}). It turns out that the Jacobian
which is constructed with this knowledge from \eqref{eq:nonergodicity_equation_fp} behaves like the
stability matrix in classical MCT, as one of its eigenvalues approaches 0 at the delocalization
transition. Similar to the schematic calculation, it is possible to derive a scale-free fixed point
iteration for the corresponding eigenvector, the so called critical amplitude (see \eqref{eq:beta:critical_amplitude_iteration} and Chapter~5.4 of
\cite{GruberPhD2019} for the details of this tedious calculation). While the critical amplitude remains finite for small wave vectors in classical MCT, (cf.~Fig.~4.6 in \cite{Goetze2009}), we find that in our model the critical amplitude diverges for small wave vectors: like $q_x^{-2}$ in the perpendicular direction and like $q_z^{-1}$ for the parallel direction (see inset in the lower panel of Fig.~\ref{fc-Aq_t100_theo}). This renders it impossible to normalize the critical amplitude by its norm. Instead, we have to fix the value of the critical amplitude at a certain wave vector.

\begin{figure}
\includegraphics[width=0.95\figurewidth]{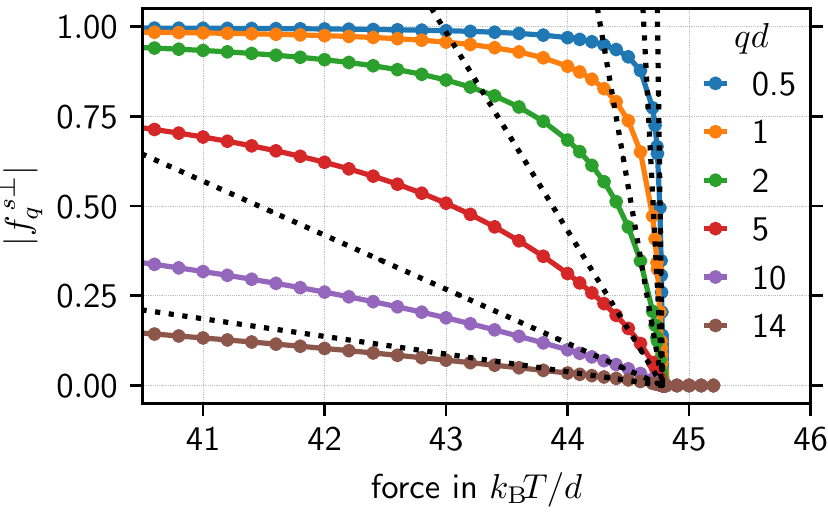}
\includegraphics[width=0.95\figurewidth]{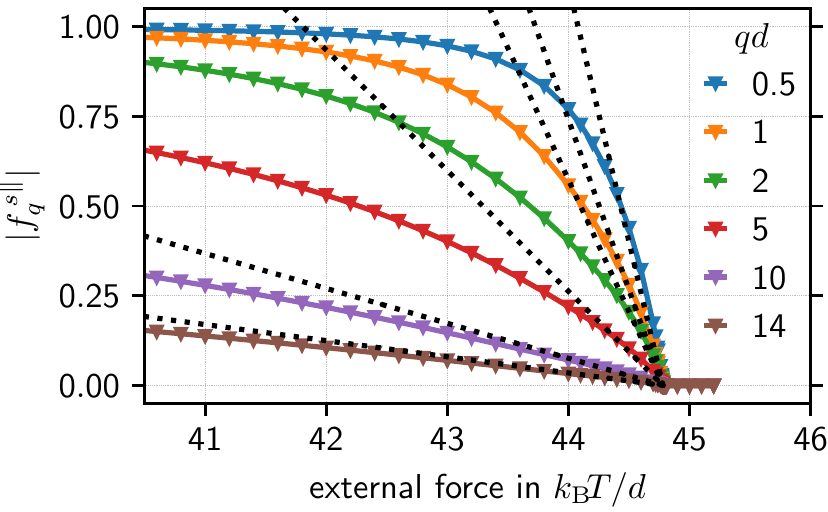}
\caption{Nonergodicity parameters $|f^s_{\vq}|=\lim_{t\to\infty}|\phi^s_{\vq}(t)|$ for
perpendicular (top panel) and parallel (bottom panel) wave vectors with $q=0.5,1,2,5,10,14$ from top
to bottom. The dashed line indicates the critical force and the dotted lines the asymptotic
behavior close to the critical force.
 \label{nep_scaling}}
\end{figure}

This divergence can be understood by checking the implications of the critical amplitude.
For example, it determines the scaling of the nonergodicity parameters close to the critical point
as shown in Fig.~\ref{nep_scaling}. For a type A transition, we have the following asymptotic
expansion for $\delta:=(\Fext-F^c)/F^c < 0$
\begin{equation}
  f^s_q = - h_q \delta + \Landau(\delta^2)
\end{equation}
with an appropriate scaling of $h_q$. The dashed lines in Fig.~\ref{nep_scaling} are given by this relationship and the values of $h_q$ are shown in the lower panel of Fig.~\ref{fc-Aq_t100_theo}. The scaling factor for $h_q$ is set at the largest wave vector in the perpendicular
direction. We notice that the decay of the nonergodicity parameter becomes steeper and steeper for
smaller wave vectors. This effect is more prominent for wave vectors perpendicular to the force
direction as was the divergence of the critical amplitude. 

For the analysis of the critical dynamics, we start out with \eqref{eq:eom_laplace} and focus on the
long time behavior around the critical force. Formal calculations show that we can expect a
$s^{-1/2}$ or a $t^{-1/2}$ power-law scaling of the correlation functions at the delocalization
transition (see Chapter~5.6 in \cite{GruberPhD2019} for the details). These findings can also be
confirmed numerically. We note, however, that the onset of this scaling depends on the wave vector
magnitude. The smaller the magnitude, the later (longer times or smaller Laplace frequencies) the
onset of the critical power law. 

This finding also disproves the existence of a master curve for the
correlation functions for all $q$ at a finite distance from the critical point, which is usually assumed for the beta-scaling analysis. Nevertheless, we find
a master curve for the primitive memory kernels, i.e.\ a factorization of the Laplace frequency and wave vector dependency. With this information, we can
work out the critical power laws for the moments, which can be traced back to the evaluation of a
combination of primitive memory functionals (see Chapter~5.8 of \cite{GruberPhD2019} for details).
We find $\left\langle z\right\rangle \propto t^{-\alpha}$, $\left\langle (x-
\mean{x})^2\right\rangle \propto t^{-\alpha}$ and $\left\langle (z-\mean{z})^2\right\rangle \propto
t^{-2\alpha}$ using a general critical power law with exponent $\alpha$ (i.e.\ $\Phi^s_q (s) \propto
s^{\alpha}$). With the choice $\alpha=-1/2$, based on the results from the schematic model, the exponents above reduce to $1/2$, $1/2$ and $1$.
%\noteAP{Why is this a choice? You can motivate it by the schematic model, can't you?} \noteMG{I can motivate this exponents by the schematic model, but I can also find different exponents for the full model or for the simulations ($\alpha=0.9$). Nevertheless, the relation between the exponents for the different moments holds in all cases.}

For the behavior close to the critical force, we can identify a timescale, which scales like $\delta^{-2}$ as can be observed in Fig.~\ref{rp-theo}. This is the same scaling as derived for the schematic model as discussed in the previous section. This highlights again that the schematic model represents the critical features of the full model. 

With the knowledge of the scaling of the memory functionals, we can also discuss the scaling of the stationary velocity as function of the distance to the critical force. The stationary velocity $v_\text{st}:=\lim_{t\to\infty} \partial_t \mean{z}(t)$ is related to the integral over the memory functional $m_0^{zz}(t):= \lim_{\vq\to 0} m_{\vq}^{zz}(t)$ via 
\begin{equation}
    v_\text{st} = \frac{\Fext}{1 + \int_0^\infty m_0^{zz}(t') \du t' }
    \label{eq:stationary_velocity}
\end{equation}
(see Eq.~(2.189) in \cite{GruberPhD2019} and \cite{Gazuz2013}). Below and at the critical force, this memory integral diverges, leading to a vanishing stationary velocity. Above the critical force, this integral becomes finite as the memory functional decays on a certain timescale. A scaling argument (see Section~5.6.3 of \cite{GruberPhD2019} and Appendix \ref{sec:scaling_memory_kernel}) suggests $\int_0^\infty m_0^{zz}(t')\du t' \propto \delta^{-1}$, which results in $v_\text{st}\propto \delta$ to first order for $\delta>0$. 

Summarizing the beta-scaling analysis for the full model, we find for the given discretization a
type A transition with a critical power law with exponent $-1/2$ for the correlation functions. The
critical amplitude can be determined via a scale-free fixed-point-equation and has diverging
values in the limit $q\to 0$. The critical power law is most pronounced for large wave numbers,
while being shadowed for small wave numbers. The mean displacement and the variances perpendicular
and parallel to the force direction exhibit critical power laws as well with exponents $1/2$,
$1/2$ and $1$, respectively.

% to derive:
% 
% critical laws: 
% 
% $\phi^s_\vq(t) = A_q t^{-1/2}$
% 
% % derive equation for critical amplitude
% % 
% % $f^s_\vq(\Fext) = c*A_q (\Fext-\Fc)$ for large $q$
% 
% $\left\langle z\right\rangle \propto t^{1/2}$, $\left\langle (x-\Delta x)^2\right\rangle \propto
% t^{1/2}$ ,$\left\langle (z-\Delta z)^2\right\rangle \propto
% t^{1}$

\section{Simulations}
\label{sect_sim}

A polydisperse system of quasi-hard spherical particles is simulated with Langevin dynamics. The
equation of motion for particle $j$ is given by \cite{Dhont1996}:
\begin{equation}
m \ddot{\vec{r}}_j = 
  -\gamma_0 \dot{\vec{r}}_j + \vec{\eta}_j(t) + \sum_i \vec{F}_{ij}
  + \delta_{j1} \vec{F}_{ext},
\end{equation}
where the friction with the solvent, $\gamma_0 \dot{\vec{r}_j}$, is given by the
coefficient $\gamma_0$, and $\vec{\eta}_j(t)$ is a random force linked to the friction coefficient via the fluctuation-dissipation theorem \cite{Dhont1996}. The particle-particle interactions, $\vec{F}_{ij}$, is given in our case by the inverse-power potential:
\begin{equation}
V(d_{ij})=\kT(d/d_{ij})^{-36}
\end{equation}
with $d_{ij}$ the center to center distance between particles. Finally, the external force
$\vec{F}_{ext}$ is applied only to the tracer, labeled as $j=1$.

\begin{figure}
\includegraphics[width=0.95\figurewidth]{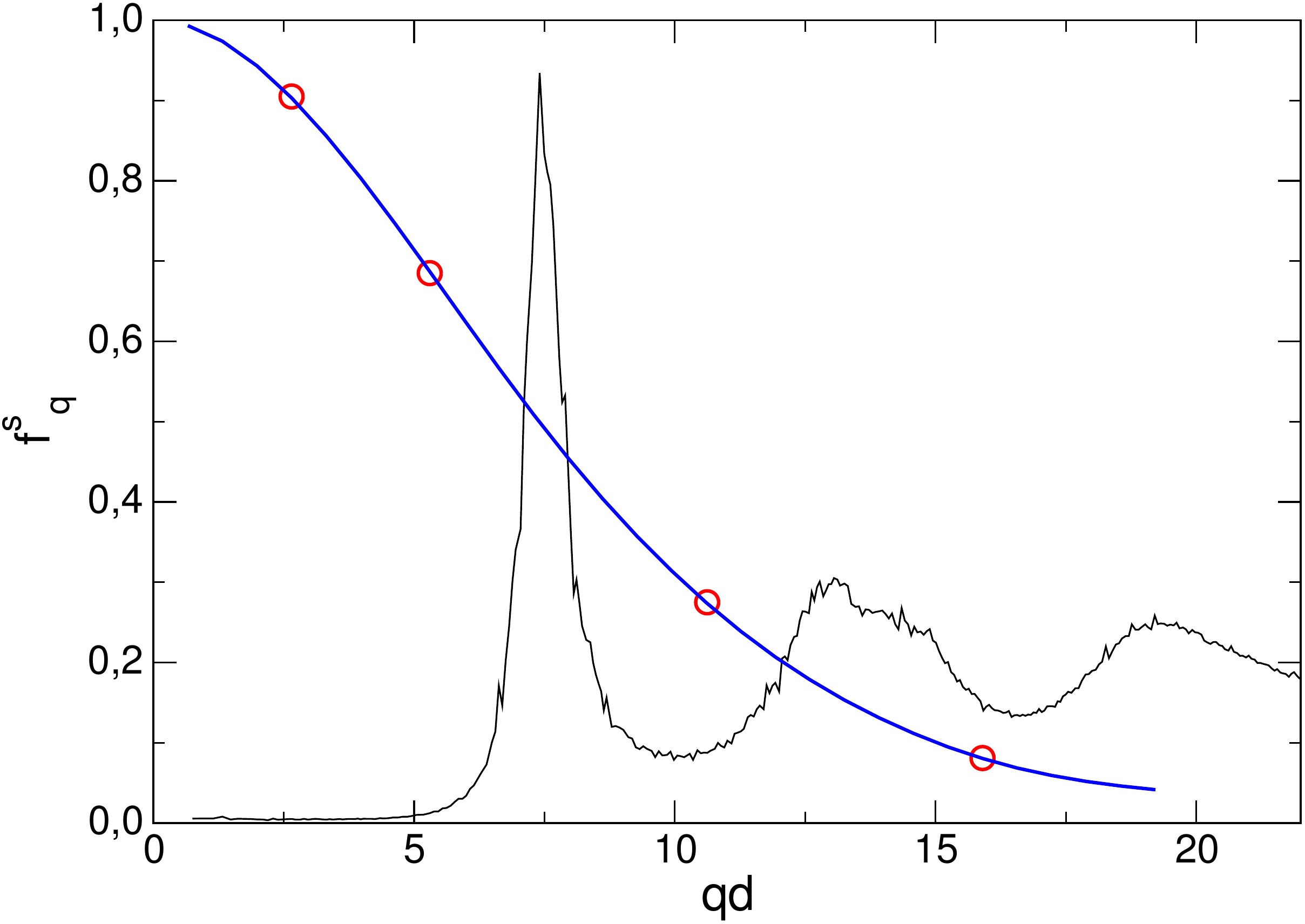}
\caption{Nonergodicity parameter and structure factor of the glassy host system. The red circles identify some typical wave vector moduli studied below. \label{bulk-glass-sims}}
\end{figure}

The simulated glass has a volume fraction, $\varphi=0.62$, calculated considering hard spheres of
diameter $d$. The glass transition for this system has been estimated previously by an MCT analysis
yielding $\varphi_g=0.596$ \cite{Voigtmann2004,Weysser2010}. Hence, this system is as far from the
glass transition as the system used for the theoretical calculations. The preparation and properties
of this glass have been discussed previously \cite{Puertas2010,Gruber2016}. In essence, the system is equilibrated in a fluid state with the same density and moderate attractions; the attractions are then suddenly removed, leaving only the
repulsive interactions. The system is then aged for a long time. The dynamics of the aged glass does
not show any sign of further evolution for the same time range used in the study of
microrheology presented here. Fig.~\ref{bulk-glass-sims} presents the nonergodicity parameter of
the glass and the structure factor, for reference. Further details of the preparation procedure can
be found in Refs. \cite{Puertas2010,Gruber2016}.

In our simulations of microrheology we focus on the transient regime. A particle is randomly
selected as tracer and at $t=0$, a constant external force is applied to it, pulling it through
the system. The tracer trajectory is monitored as a function of time. The tracer is allowed to
travel through the simulation box more than once, as we could not identify any different behaviour
between the first and consecutive passages. The results presented below are the average over ca.
5000 independent trajectories (tracers) for every force. For $\Fext=80\,kT/d$, which we identify as
the critical force, 25000 trajectories have been simulated.

The main observable of interest is the tracer position correlation function, as discussed in the
theory section, namely, the Fourier transform of the tracer displacement probability distribution.
In the simulations, the wave vector is restricted due to the periodic boundary conditions. Even more, since the dynamics of
the tracer depends on the direction, only wave vectors with moduli multiple of $2\pi/L$, are allowed
in the force direction, with $L$ the dimension of the simulation box in this direction. Some of the
wave vectors studied below (in particular in Fig.~\ref{fq_sims}) are shown as red circles in
Fig.~\ref{bulk-glass-sims}.

In the simulations we consider $N=1000$ particles in a cubic box with periodic boundary conditions.
The box size is set by the volume fraction of the particles, $\phi=0.62$, yielding $L=9.48\,d$, with $d$ the average particle diameter.  All particles have the
same mass, $m$; diameters are distributed according to a flat distribution of width $\Delta=0.1d$, to avoid crystallization.
In the simulations we set to $m=1$, $d=1$ and the thermal energy is
$\kT=1$. The solvent friction coefficient is set to $\gamma_0=10 \sqrt{m\kT}/d$, giving a diffusion
coefficient of the free particle $D_0=\kT/\gamma_0=0.1 d \sqrt{\kT/m}$. Time is measured in units of
the Brownian time $\tau_B=d^2/D_0$. The equations of motion are integrated using a Heun algorithm
\cite{Paul1995} with a time step of $0.00025 d \sqrt{m/\kT}=2.5\cdot 10^{-5} \tau_B$.

\section{Results and discussions}
\label{results}
The theoretical predictions for the critical force in the section about the bifurcation analysis
were based on the evaluation of the long-time limit of the correlation functions. This limit is,
however, unreachable in the simulations, where a finite-time analysis must be performed. We will
thus first study if the critical force and amplitude can be properly estimated from a finite time analysis. Then, the power-law behaviour of the correlation function and tracer displacement at the critical force, predicted by our model, are tested with simulations. Finally, we will show that the stationary velocity for a long but finite time provides an additional estimation of the critical force. These results will provide three different methods of determining the critical force and the properties of the transition, corroborated by our simulations, that can be used in experimental systems.

\subsection{Estimation of the critical force}
\label{sec:critical_force_estimate}
In this section, we will analyze a procedure to estimate the critical force based on finite time results. While the critical force is defined based on the vanishing of the nonergodicity parameter (the long-time-limit of the tracer position correlation function), this cannot be achieved in simulations. Therefore, we model the procedure for the simulations by evaluating the theoretical result for the tracer position correlation function at a large but finite time as an approximation to the nonergodicity parameter. From these values, the critical force and amplitude are estimated, and compared with the values obtained from the true long time limit.

\begin{figure}
\includegraphics[width=.95\figurewidth]{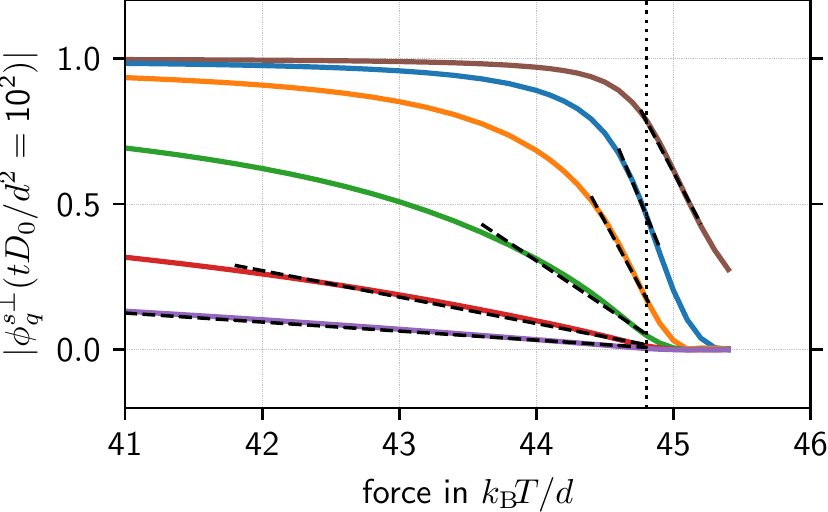}\\[1em]
\includegraphics[width=.95\figurewidth]{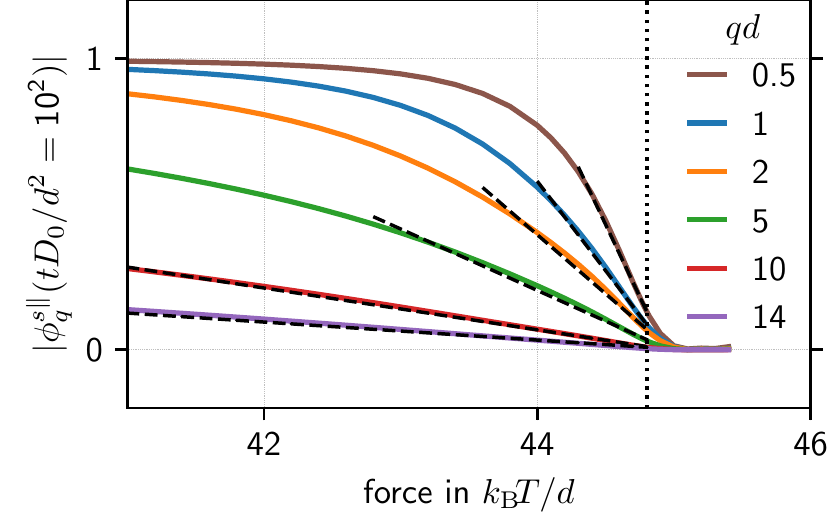}
\caption{
Values of the tracer position correlation function for $t=10^2 d^2/D_0$ as a function of the
external forces for different wave vectors perpendicular to the external force (upper panel) and
parallel to it (lower panel). Numerical solutions of the MCT equations are shown. The dotted vertical line indicates the true critical force. The dashed
black lines indicate linear fits to the nonergodicity parameters to determine the critical force
shown in Fig.~\ref{fc-Aq_t100_theo}.
\label{fq_t100_theo} }
\end{figure}

The values of the theoretical correlation functions as a function
of the external force for different wave vectors at the finite time $t D_0/d^2 = \num{E2}$ are
shown in Fig.~\ref{fq_t100_theo}. Wave vectors parallel and perpendicular
to the force direction are presented; since the correlation functions in the force direction are complex, the modulus is studied. While the type A behavior for large wave vectors
is very similar to the results for the true long-time limit (shown in Fig.~\ref{nep_scaling}), there
are some differences for small wave vectors. This graph shows that the diverging slope of $f_q$ as
a function of the external force at the critical force value is replaced by a finite slope, which
intersects with 0 at larger forces. This will induce systematic errors in the estimation of the
critical force value when analyzing small wave vectors. Although this effect is noticed for both
directions, the transversal one shows a more dramatic effect, consistent with the
theoretical analysis for the critical amplitude.

\begin{figure}
\includegraphics[width=.95\figurewidth]{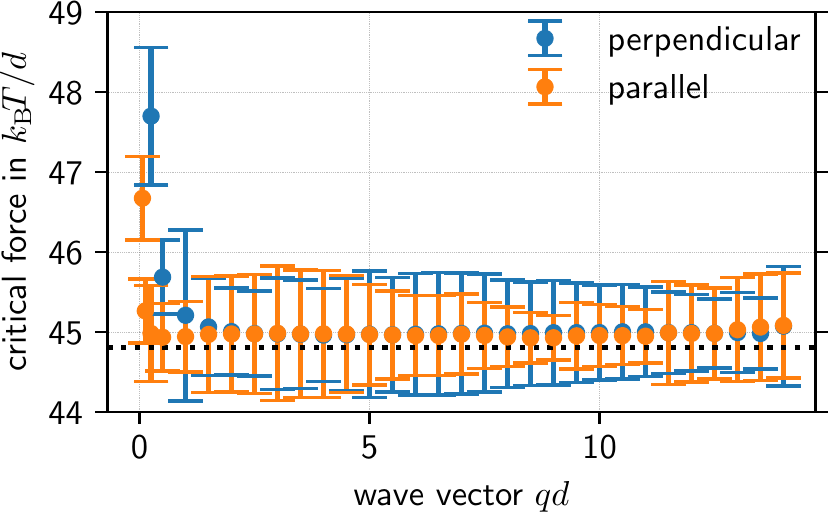}\\[1em]
\includegraphics[width=.95\figurewidth]{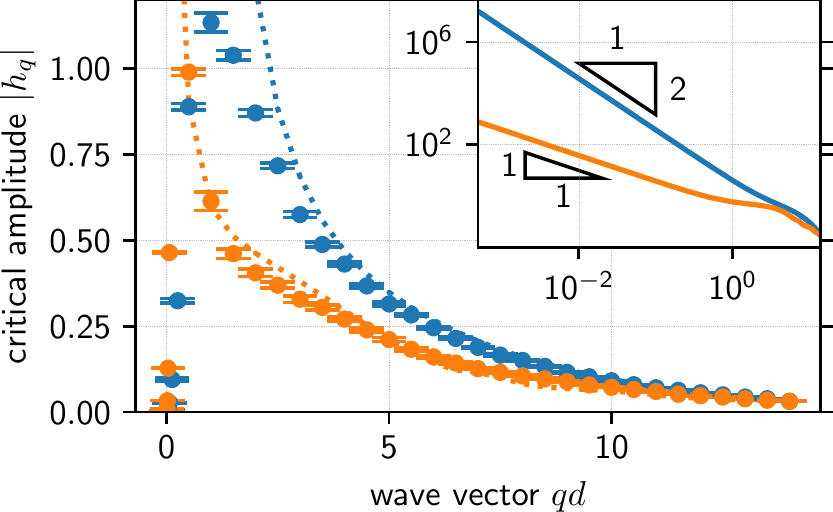}
\caption{Critical force (upper panel) and critical amplitude (lower panel) obtained from time-dependent MCT calculations as a function of the wave vector. Symbols show data  estimated from the
linear fittings of $f_c$ vs. $\Fext$  for finite times (see
Fig.~\ref{fq_t100_theo}). Errorbars indicate the uncertainties in these parameters arising from the fit. The horizontal dashed line in the upper panel marks the true critical
force and the dashed lines in the lower panel are the critical amplitudes obtained in the limit $t\to\infty$ of $\phi_q^s(t)$. The inset shows the same critical amplitudes on a log-log scale establishing the $(q_x)^{-2}$ and $(q_z)^{-1}$ divergences in the low $q$ limit.
\label{fc-Aq_t100_theo}}
\end{figure}

The analysis of the correlation function at finite times as a function of the external force is
presented in Fig.~\ref{fc-Aq_t100_theo}. Here, the value of the critical force (upper panel) is
extracted from the intersection of the linear fit of $f_q$ vs. $\Fext$ with the $x$-axis. The
range of the fit has been adjusted to capture the linear behavior best. The errorbars
indicate the uncertainty from the fit.
As mentioned previously, the critical force for small wave vectors is overestimated, more
prominently when the wave vector is perpendicular to the external force. However, for not-too-small
wave vectors, the estimate is independent of $\vq$ (direction or modulus), and more importantly, agrees with the value extracted from the analysis of
the long-time limit.

The critical amplitude (lower panel) at the finite time is given by the slope of this fit (symbols). Since we have complex valued correlators, we fit real and imaginary part separately and show the modulus of the resulting complex critical amplitude at finite times. 
It shows stronger deviations from the true critical amplitude (dashed lines), which disappear only
for large wave vectors.
Both data for the longitudinal and transversal wave vectors display a maximum as a function of $q$,
which is not observed in the analysis of the proper long time limit. Looking back at Fig.~\ref{fq_t100_theo}, we realize that this behavior arises from the fact that the correlation functions have not yet decayed to zero above the critical force at finite times. This implies that for small wave vectors,
particularly when $\vq$ is perpendicular to the force, the proper critical amplitude is underestimated. 

\begin{figure}
\includegraphics[width=.95\figurewidth]{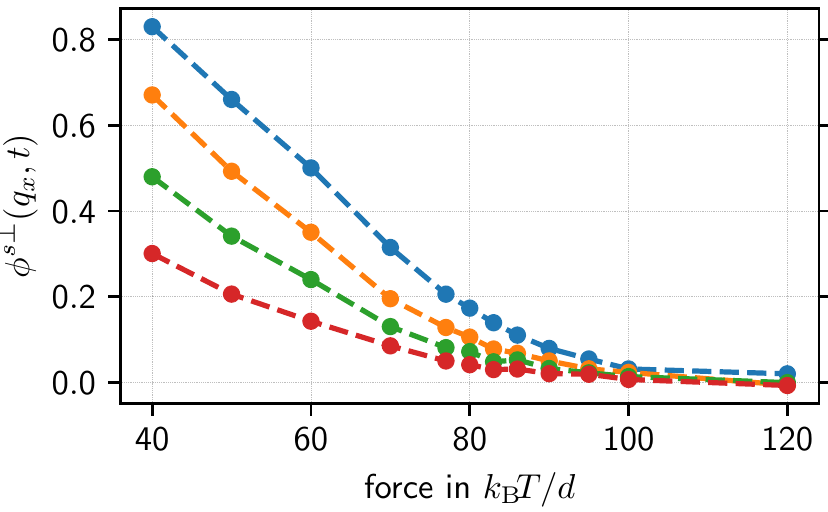}
\includegraphics[width=.95\figurewidth]{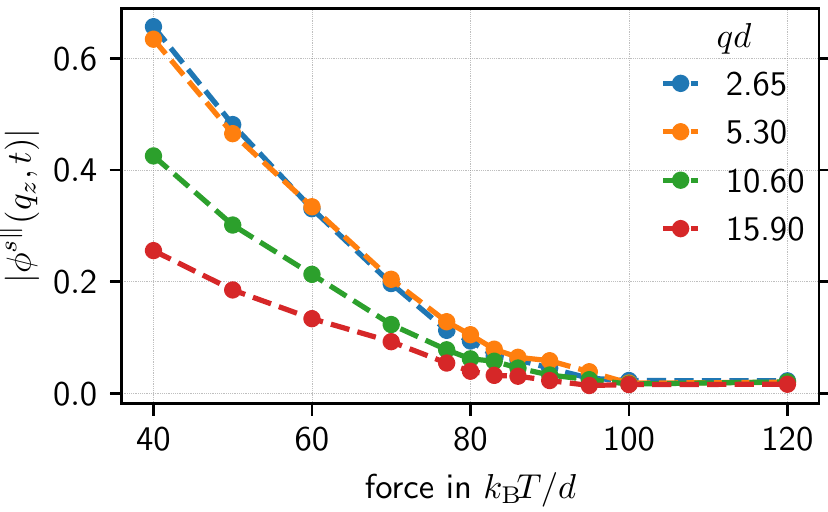}
\caption{
Value of the tracer position correlation function from simulations for different
wave vectors perpendicular to the external force (upper panel) and parallel to it (lower
panel). The wave vector moduli are given in the legend, and are marked by the red circles in
Fig.~\ref{bulk-glass-sims}. \label{fq_sims} }
\end{figure}

With these differences in mind, we now turn to the simulation data, and analyze the tracer position
correlation functions averaged over time in the range $[10, 25] d^2/D_0$. It must be remembered that
the wave vector in the analysis has to be compatible with the periodic boundary conditions.
Fig.~\ref{fq_sims} presents the value of the correlation function as a function of the external
force for different wave vectors (similar to Fig.~\ref{fq_t100_theo}, wave vectors parallel and
perpendicular to the force are considered). As predicted by the theory, the correlation function is
lower for wave vectors parallel to the external force. The anomalies reported within the theory for
small wave vectors are not observed here, because these wave vectors are not accessible in the
simulations.

\begin{figure}
\includegraphics[width=.95\figurewidth]{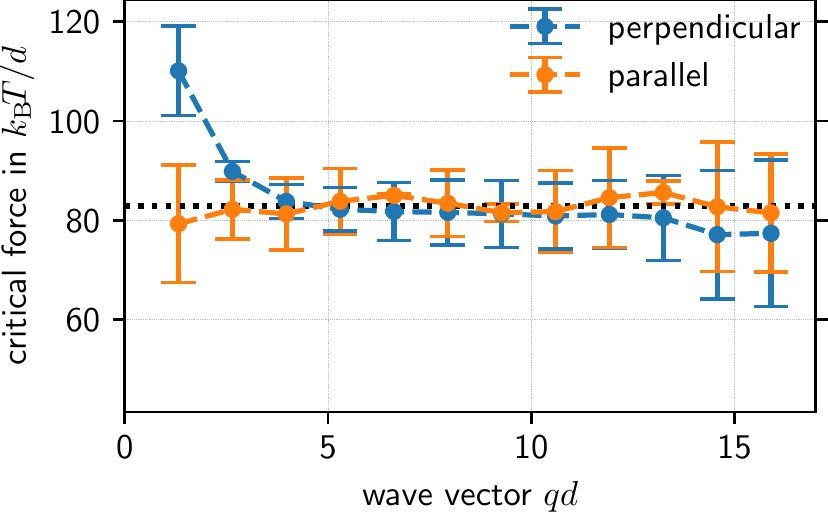}
\includegraphics[width=.95\figurewidth]{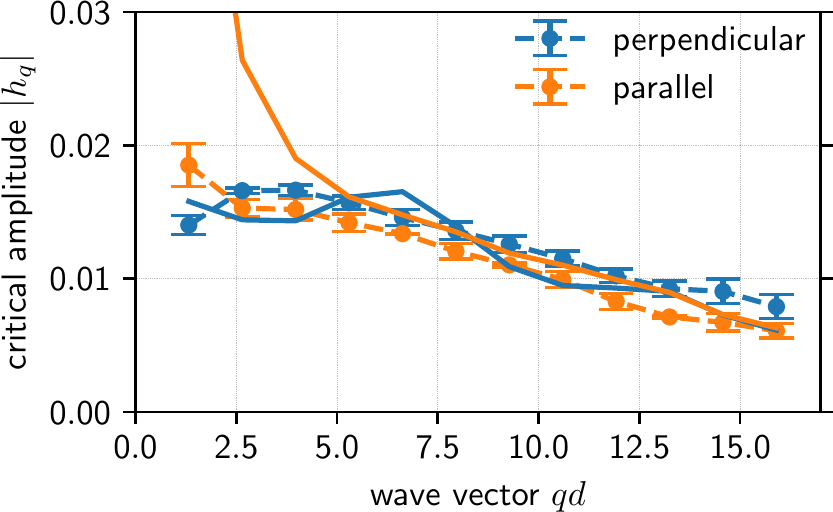}
\caption{Critical force (upper panel) and critical amplitude (lower panel) from the simulations for wave vectors parallel and perpendicular to the external force as a function of the
modulus of the wave vector. Symbols show data from the linear fitting of $f_c$
vs. $\Fext$ (see
Fig.~\ref{fq_sims}). The dashed horizontal line in the top panel indicates the average
critical force for $qd>\num{2.5}$ of \SI{82.4}{\kT/\d}. The solid  lines in the lower panel are
estimates of the critical amplitude from the long time behaviour of the tracer position correlation function for $\Fext=\SI{80}{\kT/\d}$ (see text below).
\label{fc-Aq_sims} 
}
\end{figure}

The values of these correlation functions are fitted with a linear model, following the theoretical
prediction. The resulting critical force (from the intercept of the fitting with the $x$-axis) and
critical amplitude (slope of the fitting) are shown in Fig.~\ref{fc-Aq_sims}. The critical force
does not depend on the wave vector modulus or direction, within the error bars, for large wave
vectors, but increases when it is estimated from small wave vectors, in agreement with the
theoretical results in Fig.~\ref{fc-Aq_t100_theo}. Averaging those results for
$\num{2.5}<qd<12.5$, we obtain a critical force of \SI{83.1+-5.8}{\kT/\d} from the parallel
direction and \SI{81.5+-5.8}{\kT/\d} for the perpendicular direction. Given the theoretical analysis
presented above, we therefore estimate from these results that the critical force in the
simulations is about $F_c = \SI{80}{\kT/\d}$, taking into account that the finite time analysis
overestimates the critical force. This value is significantly higher than in the theory
$F_c=\SI{44.79}{\kT/\d}$, although the glass in the theory is ideal, in contrast with the real
glass in the simulations.

The critical amplitude, plotted in the lower panel of
Fig.~\ref{fc-Aq_sims}, decreases in qualitative agreement with the theoretical results. For small
wave vectors, it shows a maximum when the wave vector is perpendicular to the external force, also in
agreement with the theoretical results of Fig.~\ref{fc-Aq_t100_theo}. However, other details of the
theoretical analysis are absent in the simulations; in particular, the critical amplitude does not
become independent on the wave vector direction, and when the wave vector is parallel to the
external force the simulation data does not show a maximum.

When the wave vector is parallel to the external force, it must be recalled that the tracer position
correlation function is complex valued (thus, the modulus of the correlation function has been
studied) \cite{Gazuz2009,Gruber2016}. For increasing force, the real part of the nonergodicity
parameter becomes negative, while the imaginary part describes a maximum. Both components tend
linearly to zero at the critical force; the modulus is dominated by the imaginary part. The analysis
of the correlation functions at a finite time (not shown) yield the same conclusions as drawn
previously. The comparison with the simulations, however, is more difficult due to the large
noise-to-signal ratio in the real part.

Summarizing this section, we can identify the critical force consistently from linear extrapolation of the tracer position correlation function at long, but finite times in theory as well as in simulations. The uncertainty for the values of the critical force is largest for small wave vectors due to their strong variation around the critical force. We can also extract the critical amplitudes, which increase with decreasing wave vector, for both, simulations and theory. 

\subsection{Long time behaviour of the tracer position correlation function}
\begin{figure*}
\includegraphics[width=1.95\figurewidth]{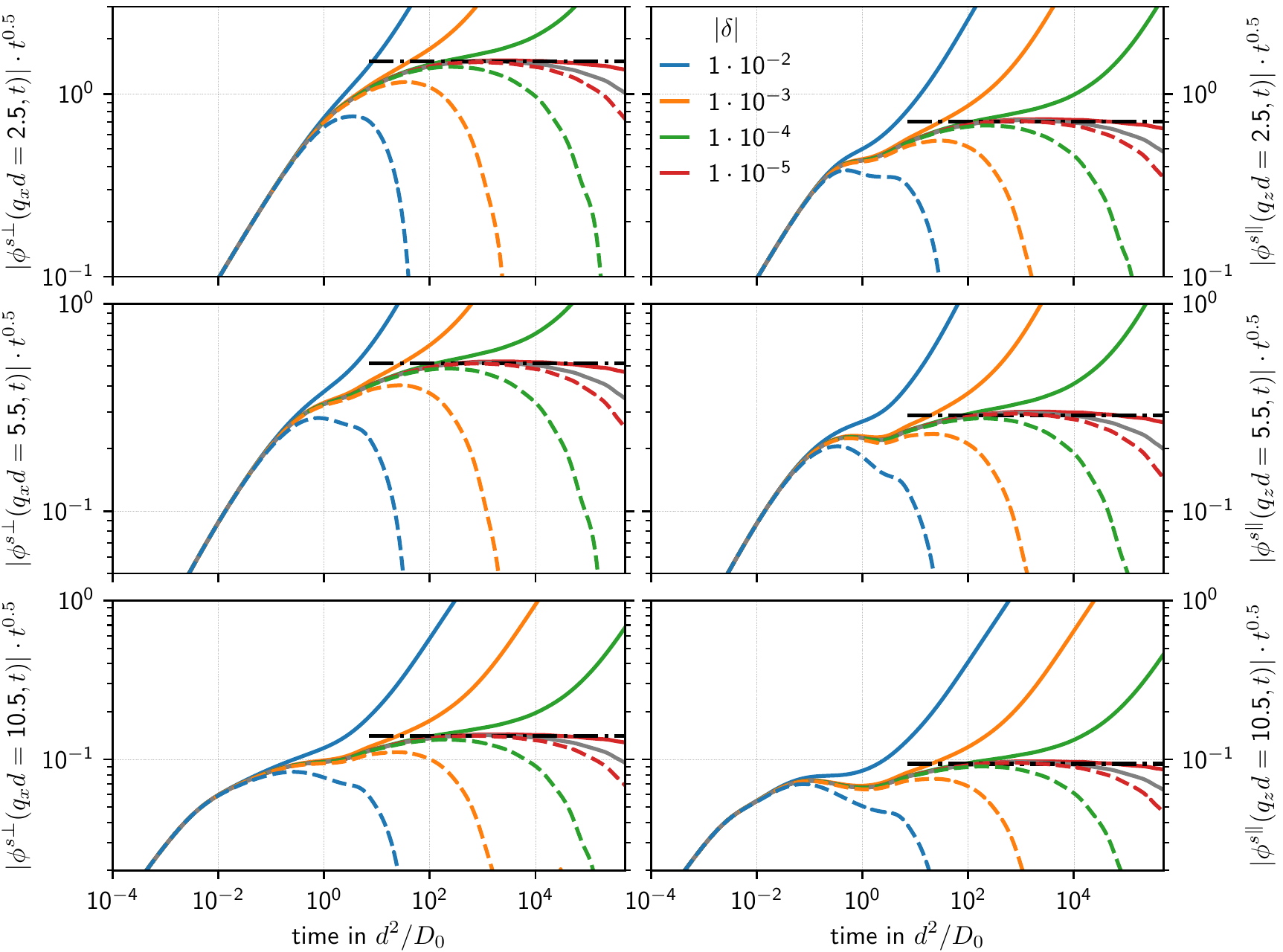}
\caption{
Rectification plots for the tracer position correlation function for wave vectors perpendicular (left
panels) and parallel to the external force (right panels) from the theory. Several forces around the
critical one ($\delta = (\Fext-F^c)/F_c$) are presented. Dashed lines indicate forces above the
critical force, solid lines forces below. The solid grey line indicates the critical
behavior. Different wave vectors are presented:
$qd=2.5$ (top panels), $qd=5.5$ (intermediate panels) and $qd=10.5$ (bottom panels). \label{rp-theo} }
\end{figure*}

We analyze in more detail the tracer position correlation functions in the vicinity of the critical
force estimated previously. The theory predicts a power law decay with exponent $-1/2$. The onset of
this power law depends on the wave vector and its prefactor is described by the critical
amplitude for large wave vectors. For a better resolution of the critical force, we use for this
section $N_\text{log}=25$ points in the nonuniform part of the grid for the theoretical
calculations as described in the section about the numerical details. 

Fig.~\ref{rp-theo} presents rectification plots of the modulus of the tracer position correlation
function with the predicted behaviour. While forces well above or below the critical one do not show
the power-law decay, forces close to that indeed follow the expected trend, for all wave vectors
moduli and directions. Note that the start of this
power-law behavior occurs for earlier times the larger the wave vectors are. It should be also
stressed that the trend is also followed even for small wave vectors, despite the difficulties in
handling low wave vectors in the theory.
Noteworthy, 
%the correlation functions reach the power-law decay for longer times the lower the wave vector, but also depending on the wave vector direction; namely, 
the critical behaviour is more difficult to be observed for small wave vectors perpendicular to the external force.

% \begin{widetext}

\begin{figure*}
\includegraphics[width=1.95\figurewidth]{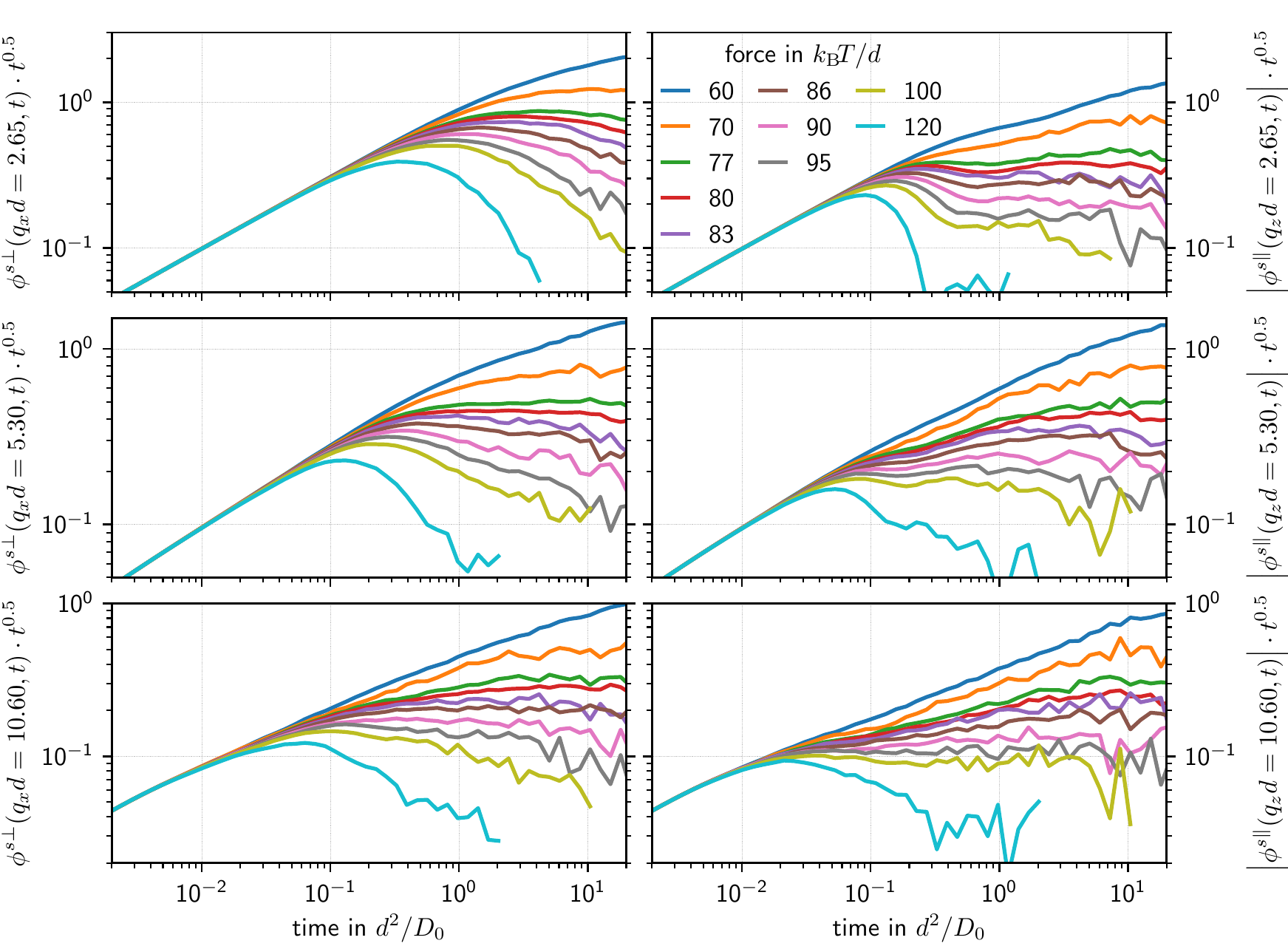}
\caption{ 
Rectification plots for the tracer position correlation function for wave vectors perpendicular
(left panels) and parallel to the external force (right panels) from simulations. Several forces
around the critical one are presented; from top to bottom: $\Fext=60$, $70$, $77$, $80$ (red line),
$83$, $86$, $90$, $95$, $100$ and $\SI{120}{\kT/\d}$. Different wave vectors are presented:
$qd=2.65$ (top panels), $qd=5.30$ (intermediate panels) and $qd=10.60$ (bottom panels). 
\label{rp-sims} }
\end{figure*}

% \end{widetext}

As discussed previously, when the wave vector is parallel to the external force the tracer position
correlation function is complex valued. The real part takes on negative values for large wave
vectors and tends to zero from below, while the imaginary part is always positive and describes a
maximum. Both components indeed exhibit the $t^{-1/2}$ decaying to zero (not shown). 

The same rectification plots for the simulation data are shown in Fig.~\ref{rp-sims}. In fact, the
critical force identified by our previous analysis, $\Fext \approx \SI{80}{\kT/\d}$, shows a
plateau, except for the lowest wave vector studied in the perpendicular direction. In agreement
with the theoretical expectations, forces well below the critical force deviate upwards, while
forces above it deviate downwards. Also, it is observed that the critical behaviour is reached for
longer times the smaller the wave vector. This justifies the absence of a power-law behaviour in the
top-left panel.

Note that the observation of a power-law decay in the tracer position correlation function at long
times for all wave vector moduli and directions can be used as an alternative criterion to identify
the critical force. As observed in Fig.~\ref{rp-sims}, this yields the same result as the fitting of
the nonergodicity parameters with the external force, $F_c\approx \SI{80}{\kT/\d}$. However, because
the observation of a specific behaviour in the correlation function is arguable, the extrapolation of
the nonergodicity parameters appears as a more robust method.

The prefactor of the $t^{-1/2}$ behaviour, given by the value of the long time plateaus in
Figs.~\ref{rp-theo} and \ref{rp-sims}, is the critical amplitude, up to a constant factor arising
from a microscopic timescale. According to the theory, this should agree with the slope of the
nonergodicity vs. the external force (up to this constant factor). The critical amplitude for the
theory, is plotted in the lower panel of Fig.~\ref{fc-Aq_t100_theo} as dashed lines; to avoid the
ambiguity of the microscopic timescale, both sets are matched at the highest wave vector in
the perpendicular direction. At large wave vectors, both estimations of the critical amplitude
decay monotonously, and agree for the highest wave vectors. However, differences appear for small
wave vectors, that become qualitative for smaller wave vectors as the slope of $\phi^s(t)$
vs. $\Fext$ describes a maximum (explained in Sec.~\ref{sec:critical_force_estimate}), while the true critical amplitude grows continuously with decreasing $q$.

In the simulations, the prefactor of the $t^{-1/2}$ behaviour in the tracer position correlation
function has been estimated by averaging the value of $t^{1/2} \phi_q(t)$ in the interval $t \in
[10, 25] d^2/D_0$. The results are presented again in comparison with the estimation from the
long-time value of $\phi^s(t)$ vs. $\Fext$ in Fig.~\ref{fc-Aq_sims}. As in the previous case, the
data have been matched at the largest wave vector in the perpendicular direction. Again, the overall 
behaviour is similar between both estimates of the critical amplitude for large wave vectors, but
differences appear at small ones. % More importantly, t
The estimation from $t^{1/2} \phi_q(t)$ shows a
modulation with $q$, which is not observed in the theory or in the other estimation of the critical
amplitude, but which is almost within the error bars of the data.
The difference in magnitude between the critical amplitudes obtained from theory and simulations
can be related to the observation that the transition between localized and delocalized behavior is
occurs in a much more narrow force range in the theory than in the simulations. Hence, the slopes
are steeper and the critical amplitude is larger. 

In summary, we find a critical power-law decay with exponent $-1/2$ for the tracer position correlation function at long times in theory and simulations. This behavior is screened at small wave vectors. The force associated with the critical behavior is consistent with the critical force obtained in the previous section. Again, 

\subsection{Average tracer displacement}

Both, the analysis of the simulations and theory at a finite time show that the region of small wave
vectors is problematic because the tracer position correlation function decays very slowly. In the
simulations, this is an important problem, as the simulation time is always finite, restricting the
wave vectors that can be analyzed. In the theory, on the other hand, this regime has to be handled
with care as well to ensure that both \textit{(i)} the long time limit (either zero in the
delocalized state or finite in the localized one) and \textit{(ii)} the small wave vector behaviour
are correctly treated in the numerical integration scheme, because all modes are coupled.
Since the average tracer displacement and mean squared displacement (MSD)
perpendicular and parallel to the external force, are obtained from the small-wave-vector limit of
the correlation function, this problem prevents us from analyzing the results of the model for
these quantities.

\begin{figure*}
\includegraphics[width=1.95\figurewidth]{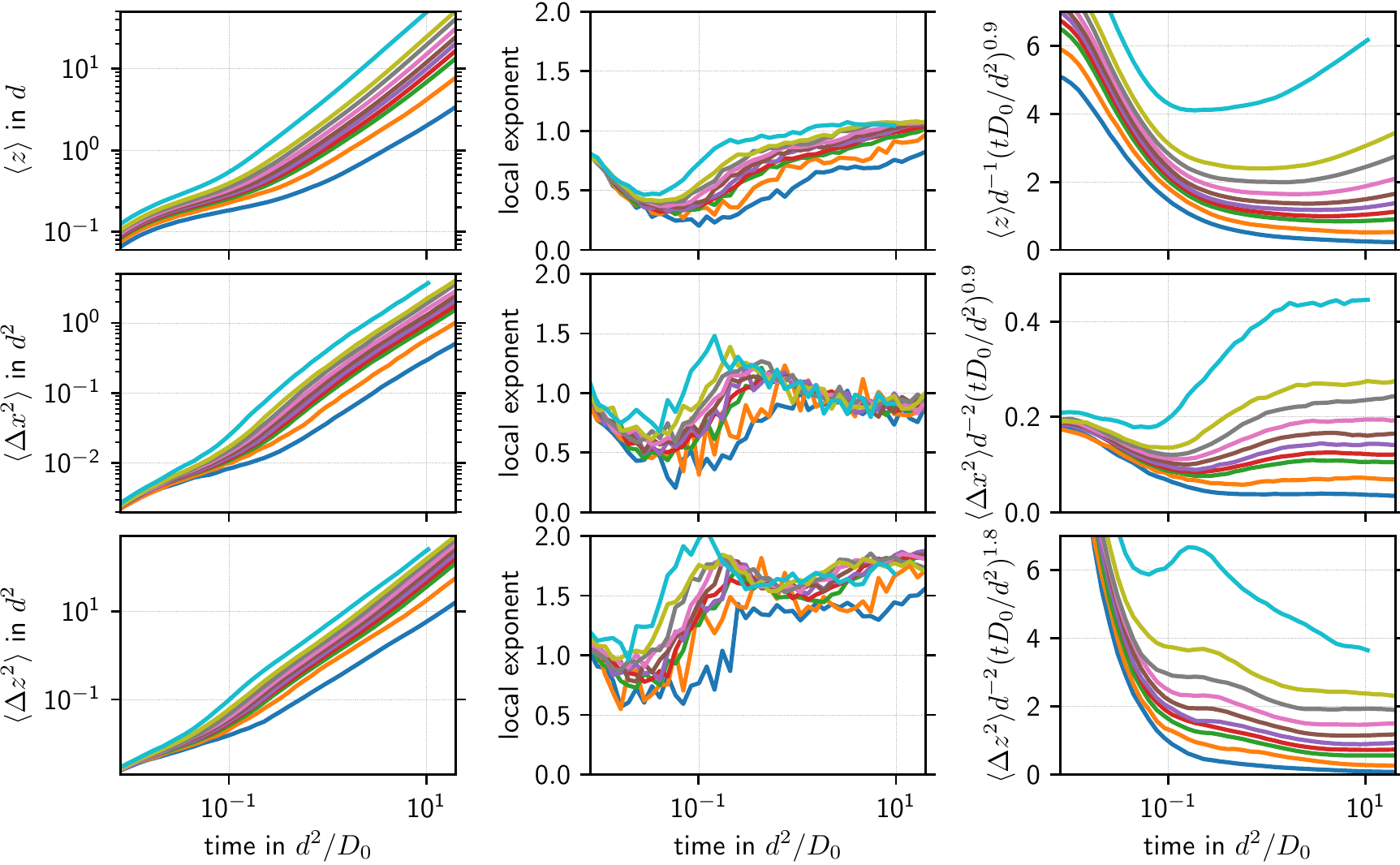}
  \caption{Critical laws for mean displacement (top row), variance $\mean{x^2}$
  perpendicular to the force (center row) and variance $\mean{\Delta z^2}$
  parallel to the force (bottom row) in the simulations for different forces
  (color code as in Fig.~\ref{rp-sims}).
  The left column shows the evolution, the center row their local exponent and
  the right row the rectified moments scaled with exponent 0.9 for the
  mean displacement and the perpendicular variance and 1.8 for the 
  variance parallel to the force.}
\label{deltaz-sims}
\end{figure*}

%\begin{figure}
%\psfig{file=figures/deltaz-sims_a.eps,width=0.8\figurewidth}
%\psfig{file=figures/deltaz-sims_b.eps,width=0.85\figurewidth}
%\caption{ \noteMG{To be replaced by Fig.~\ref{deltaz-sims}? }
%Rectification plots for the average tracer displacement (upper panel), and mean squared direction in
%the force direction  and perpendicular to it. Note that the exponents used for the rectification
%plots do not agree with the predictions from the model. The same forces and colour code as in
%Fig.~\ref{rp-sims} are used. \noteMG{how do we present the deviations from the predictions?
%Rectification plots with the predicted exponents or rectification plots with arbitrary exponents,
%which make the line for $F=80kT/d$ flat for long times?} }
%\end{figure}

The present model predicts, based on the results for the schematic model, that the tracer displacement grows according to a power law with exponent
$1/2$, $\langle \delta z\rangle \sim t^{1/2}$, for the critical force, separating 
the localized regime (where $\langle \delta z\rangle$ approaches a constant at long times)
from the delocalized regime (where $\langle \delta z\rangle$ grows linearly). For the MSD in
the direction perpendicular to the force, a similar power law is found, while in the parallel
direction, it is predicted that $\langle \Delta z^2\rangle$ grows
linearly, for the critical force.

These predictions for the tracer displacements cannot be confirmed in the simulations, as shown by
the rectification plots in Fig.~\ref{deltaz-sims}. The average tracer displacement grows faster than
the predicted critical behaviour for all forces, and in particular, at the critical force, it grows
approximately linearly (for smaller forces it grows sublinear). Similarly, the mean squared
displacement both in the parallel and perpendicular directions do not follow the predicted
behaviour for the critical force (intermediate and bottom panels of Fig.~\ref{deltaz-sims}). 
The MSD in the perpendicular plane grows approximately linearly with time around and above the
critical force, while in the force direction, superdiffusion is observed (i.e. $\langle \Delta
z^2\rangle \sim t^{-2\alpha}$ with $-2\alpha > 1$) for forces above the critical one. A close look at
the local exponents (center column in Fig.~\ref{deltaz-sims}) suggests an exponent of \num{.9} for
the mean displacement and the variance perpendicular to the force and an exponent of \num{1.8} for
the variance in the force direction. Even though these absolute values do not agree with the predictions
from theory, their ratios follow the prediction that the exponent of the power law of the variance
in the force direction should be twice as large as for the two other quantities.  

To summarize this section, it is difficult to pinpoint the critical force based on the critical scaling laws for the moments. First, there is little variation in the local exponent for different forces. Second, the exponents do not coincide with the theoretical prediction, but are slightly larger. Nevertheless, the relation between the exponents for the different moments matches the theoretical prediction.

\subsection{Stationary velocity}
\begin{figure}
\includegraphics[width=0.95\figurewidth]{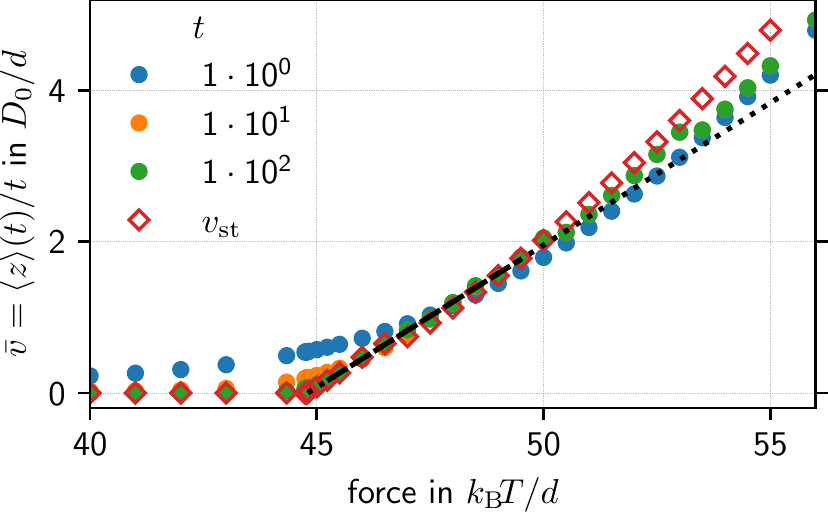}
  \caption{Average velocity for the theoretical calculations for different waiting times $t$ (circles). The stationary velocity $v_\text{st}$ (open diamond) is calculated from the memory relation \eqref{eq:stationary_velocity}. The dashed line is a linear fit to the data for $t=\num{E2}$ in the range $45<\Fext d/\kT < 50$, which is extrapolated the full range (dotted). This result was obtained using the grid with 25 points in the nonuniform part.}
\label{velocity-theo}
\end{figure}

In previous works, the average velocity was analyzed \cite{Habdas2004}, given by $\bar{v}=\Delta z / \Delta t$, where $\Delta z$ is the displacement of the probe at the end of the experiment and $\Delta t$ its duration. For long times, this converges to the definition of the stationary velocity $v_\text{st}$ as introduced in the paragraph before Eq.~\eqref{eq:stationary_velocity}. To assess the effects of finite waiting times we calculate $\mean{z}(t)/t$ in Fig.~\ref{velocity-theo} for different values of $t$ from our theory. We observe that the final velocity is overestimated below the critical force and underestimated above the critical force. This smears out the predicted linear behavior $v_\text{st}\propto \delta$ for $\delta > 0$. For $t>\num{E2}$ this linear relation is nicely visible as indicated by the dashed/dotted line. This yields another criterion to identify a lower limit for the critical force. As the velocity is underestimated for forces above the critical force, we will always obtain a value for the x-intercept which is smaller than the true critical force. 

Furthermore, we also test the prediction for the stationary velocity based on the memory integral \eqref{eq:stationary_velocity}. These results are shown as open diamonds in Fig.~\ref{velocity-theo}. They exhibit the same predicted behavior close to the  critical force, but deviate for larger forces. The differences can be attributed to numerical artifacts in connection with the stability of the solution algorithm (see Chapter 3 of \cite{GruberPhD2019} for details).

\begin{figure}
\includegraphics[width=0.95\figurewidth]{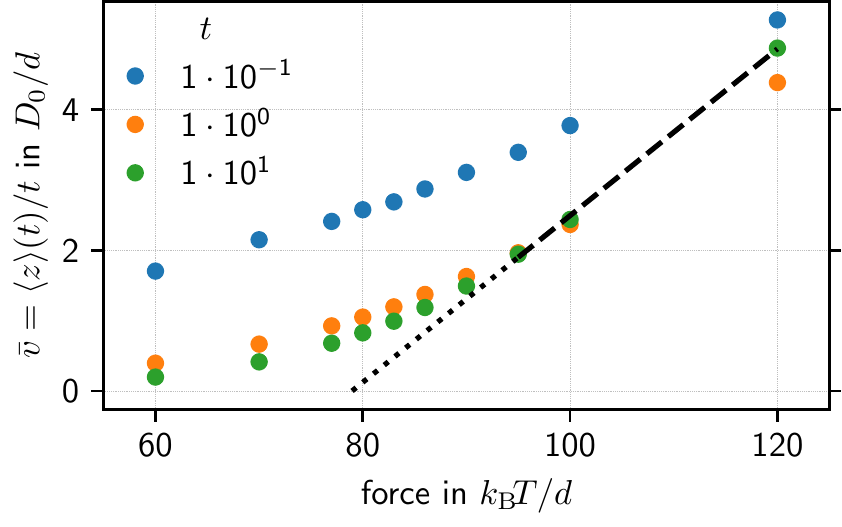}
  \caption{Average velocity for the simulations  for different waiting times $t$ (circles). The dashed line is a linear fit to the data for $t=\num{E2}$ and $85<\Fext d/\kT$, which is extrapolated the full range (dotted).}
\label{velocity-sim}
\end{figure}
The average velocity from the simulations is analyzed in Fig.~\ref{velocity-sim}. Again, we test for finite time effects (circles) and find a good agreement for $t=\num{E0}$ and $t=\num{E1}$. The average velocities cross at a force of about \SI{90}{\kT/\d} similar to the results of the theory. Above this crossing, we can identify a linear behavior. A linear fit of these data gives a lower bound for the critical force of about $\Fc = \SI{79+-3}{\kT/\d}$, which is again compatible with the previous results. As before, the range in which the velocity increases is much broader as in the theoretical calculations and the transition is less sharp than in theory. 

Concluding this section, we find that the critical force can be estimated by extrapolating the average velocity linearly to zero, when the duration of the experiment is long enough. 

\section{Conclusions}
Active microrheology provides access to the microscopic properties of a complex fluid by the action of a microscopic stress, typically exerted onto a single colloidal bead. When the host system is beyond the glass transition, a finite stress must be applied to make the system flow, or in microrheology, to delocalize the tracer. The properties of this delocalization transition are studied here for a glass of colloidal hard spheres with theory and simulations. The theory model is based on the Smoluchowski equation, where the glass is described with mode-coupling theory, and a schematic model is also provided which contains only three correlators, dropping the dependence on the wave vector but retaining the direction dependence, and with a simplified memory kernel. The model predicts that the tracer is indeed localized for small forces by the cage of its neighbours, but it can break free above a threshold force, which depends on the couplings with the bath and external force. This delocalization transition has the properties of an A-type transition within mode-coupling theory, making it difficult to observe in simulations and experiments, but also tricky in the numerical solution of the theory equations. However, some properties of the transition offer consistency checks, that have here been tested with Langevin dynamics simulations. 

The theory predictions indicate that, in the localized regime but close to the delocalization transition, the nonergodicity parameter decays linearly with the pulling force, reaching the zero-axis at the critical force. There, the tracer position correlation function decays with the square root of time, with an amplitude that is proportional to the slope of the nonergodicity parameter vs. force. While these results are based on the assumption that the time is large enough, we confirm that this predictions can also be tested when the values at a finite time are taken. Concomitant power-laws follow for the average displacement and mean squared displacement, which are observed in the simulations, albeit with different exponents.

As a result, we confirm three methods to determine the critical force for the delocalization transition. The first method is based on linearly extrapolating the values of the tracer position correlation functions at long times as a function of force. The second method uses the appearance of a critical power law with exponent $1/2$ close to the critical force. The last method is based on a linear extrapolation of the average velocity of the probe. The first two methods yield the same critical force if the
wave vectors used are not too small. For small wave vectors we find some anomalies, like the divergence of the critical amplitude and a shadowing of the critical power law. The last method requires sufficiently long experimental times. While the predicted behavior for all three methods is applicable only in a very narrow force window in the theoretical calculations, we find these signatures in the simulations over a much broader window of forces, which makes them even more suitable for applications in experimental systems. 

The critical dynamics we derived and tested holds universally  at a depinning transition from local cages as described by mode-coupling theory. At the considered delocalization transition, the vanishing of the arrested component $f^s_q$ is accompanied by the divergence of the critical amplitude for small wavevectors, see Fig.~\ref{fc-Aq_t100_theo}. Considering the orientational motion of an anisotropic particle pinned in a glass host, a finite critical amplitude can be anticipated \cite{Franosch1997}. Thus testing orientational microrheology, where an external torque is applied to an elongated probe particle in a glassy host would ideally be suited to test the predicted critical dynamics.     

\section{Acknowledgements}
We thank Gustavo Abade for his initial implementation of the algorithm to solve the dynamics of the theory. MG and MF acknowledge support from project B5 of SFB1214. AMP acknowledges financial support from the Spanish Ministerio de Ciencia under project no. PGC2018-101555-B-I00.

\bibliography{draft}

\appendix
\section{Abbreviations for the schematic model}
In this section, we give the abbreviations for the beta-scaling analysis of the schematic model:
\begin{subequations}
\begin{equation}
V_{\parallel}^{\mathrm{re}} = \frac{f_b}{\Fext^2\kappa_{\parallel}^2 + 1}
\begin{pmatrix}v_1^s\\\Fext\kappa_{\parallel} v_1^s\\ 
v_2^s \end{pmatrix},
\end{equation}
\begin{equation}
V_{\parallel}^{\mathrm{im}} = \frac{f_b}{\Fext^2\kappa_{\parallel}^2 + 1}
\begin{pmatrix}\Fext\kappa_{\parallel} v_1^s \\ - v_1^s
\\ \Fext\kappa_{\parallel} v_2^s \end{pmatrix},
\end{equation}
\begin{equation}
W_{\perp} = f_b^2 \begin{pmatrix} 
				(v_2^s)^2 & 0 & \frac{v_1^s v_2^s}{2} \\
				0 & (v_2^s)^2 & 0 \\
				\frac{v_1^s v_2^s}{2} & 0 & 0 
\end{pmatrix},
\end{equation}
\begin{equation}
 V_{\perp} = \tau_3 f_b \begin{pmatrix} v_2^s\\ - \kappa_{\perp} \Fext  v_2^s \\ v_1^s  \end{pmatrix},
\end{equation}
\begin{equation}
 U_{\perp} = v_2^s f_b \begin{pmatrix} 1 \\ \kappa_{\perp} \Fext \\ 0  \end{pmatrix},
\end{equation}
\end{subequations}

\section{Critical force of the schematic model}
\label{sec:critical_force}
In this section, we will show how to derive an analytical expression for the
critical force and the critical amplitude of the schematic model defined in Eqns.~\eqref{eq:schematic:eom}-\eqref{eq:schematic:memory_perp}
for the case $\kappa_{\parallel}=\kappa_{\perp}=:\kappa$.
The starting point is the iteration equations for the critical amplitude
\eqref{eq:schematic:fixed_point_eq_hcrit}. For the sake of simplicity, we replace $v_1^s$
by $\beta v_2^s$. The iteration equations then read
\begin{subequations}
\label{eq:fixed_point_hcrit_general}
\begin{align}
 h_1 &= \frac{f v_2^s}{\Fext^2\kappa^2 + 1} \left(\Fext \kappa \beta h_2 + \beta
 h_1 + h_3\right), \label{eq:fixed_point_h1_general}\\
 h_2 &= \frac{f v_2^s}{\Fext^2\kappa^2 + 1} \left(\Fext \kappa \left(\beta
 h_1 + h_3\right) - \beta h_2\right),
\label{eq:fixed_point_h2_general}\\
 h_3 &= f v_2^s \frac{\beta h_1 h_3 + h_1^2 + h_2^2}{\Fext \kappa h_2 + h_1} .
 \label{eq:fixed_point_h3_general}
\end{align}
\end{subequations}

First, we want to rewrite the denominator in \eqref{eq:fixed_point_h3_general}. By combining
\eqref{eq:fixed_point_h1_general} and \eqref{eq:fixed_point_h2_general}, we find
\begin{equation}
 h_1 + \Fext \kappa h_2 = f v_2^s (\beta h_1 + h_3).
\end{equation}
Substituting this back into \eqref{eq:fixed_point_h3_general}, we get
\begin{equation}
h_3 = \frac{\beta h_1 h_3 + h_1^2 + h_2^2}{\beta h_1 + h_3} \Leftrightarrow
h_3^2 = h_1^2 + h_2^2.
\label{eq:h32_h12_h22}
\end{equation}
Using again \eqref{eq:fixed_point_h1_general} and
\eqref{eq:fixed_point_h2_general} we can find an expression for the right hand
side
\begin{equation}
 h_1^2 + h_2^2 = \frac{\left(f v_2^s\right)^2}{\Fext^2 \kappa^2 + 1} \left(\beta^2 (h_1^2 +
 h_2^2) + 2\beta h_1 h_3 + h_3^2\right).
\end{equation}
Solving this equation for $h_1^2 + h_2^2$, we get
\begin{equation}
 h_1^2 + h_2^2 = \frac{\left(f v_2^s\right)^2 \left(2\beta h_1 h_3 + h_3^2\right)}{\Fext^2\kappa^2
 + 1 - \left(f v_2^s\right)^2 \beta^2}. 
 \label{eq:h12_h22}
\end{equation}
Now this equation is substituted into \eqref{eq:h32_h12_h22} leading to 
\begin{equation}
 h_3^2 = \frac{\left(f v_2^s\right)^2 \left(2\beta h_1 h_3 + h_3^2\right)}{\Fext^2\kappa^2
 + 1 - \left(f v_2^s\right)^2 \beta^2}.
\end{equation}
Here, we have eliminated the variable $h_2$. In the next step we make use of the fact, that the
iteration equation is scale invariant, i.e.\ any multiple of a solution is again
a solution. This allows us to fix $h_3=1$ and we obtain a linear equation for $h_1$. Its solution is
\begin{equation}
h_1 = \frac{\Fext^2\kappa^2 +1}{2\beta \left(f v_2^s\right)^2} - \frac{1 + \beta^2}{2\beta}.
\label{eq:h1_sol}
\end{equation}
Inserting this solution together with $h_3=1$ into \eqref{eq:fixed_point_h1_general} and
\eqref{eq:fixed_point_h2_general}, we get two equations to be solved for $h_2$
\begin{subequations}
\begin{align}
 h_2 &= \bigg(\Fext^4\kappa^4
 + \left(f v_2^s\right)^3\beta(\beta^2-1) - 1 
 \notag\\
 &\qquad - (\Fext^2\kappa^2 + 1)\left(\left(f v_2^s\right)^2 (\beta^2+1) +
 \beta f v_2^s - 2\right) \bigg) \notag\\
 &\quad \cdot \left(2\Fext\kappa\beta^2\left(fv_2^s\right)^3\right)^{-1}, 
 \label{eq:h2_sol1} \\
 h_2 &= \frac{\Fext\kappa}{2 f v_2^s} \frac{\Fext^2\kappa^2 + 1 + \left(f
 v_2^s\right)^2\left(1 - \beta^2\right)}{\Fext^2\kappa^2 + \beta f v_2^s + 1}.
 \label{eq:h2_sol2}
\end{align}
\end{subequations}

Since both equations must give the same result at the critical force, we can obtain an equation for
the critical force by equating these two equations:
\begin{equation}
\begin{split}
 0 &= \Fext^6\kappa^6 + \Fext^4\kappa^4\left(3 -\left(f v_2^s\right)^2(1 +
 2\beta^2)\right) \\
 &+ \Fext^2\kappa^2 \left( \left(f v_2^s\right)^4(\beta^4-\beta^2) 
 -2 \beta \left(f v_2^s\right)^3 \right. \\
 &\qquad\qquad\quad \left.- \left(f v_2^s\right)^2(4\beta^2 + 2) +
 3\right) \\ 
 &+ \left(f  v_2^s\right)^4(\beta^4 - \beta^2) -2\beta \left(f v_2^s
 \right)^3 - \left(f  v_2^s\right)^2(2\beta^2+1) + 1.
 \end{split}
\end{equation}
This is a third order polynomial in $\Fext^2\kappa^2$, which has the following
three solutions
\begin{equation}
\left\{ -1, \left(f v_2^s\right)^2\left(\beta^2 + \frac{1}{2}\right) - 1 \pm
 \frac{\left(f v_2^s\right)^2}{2}\sqrt{8\beta^2 + 1 + \frac{8\beta}{f v_2^s}}\right\} .
\end{equation}
As we want the critical force to be real, we have to choose the largest (positive) solution, which
is
\begin{equation}
 \Fc = \frac{1}{\kappa}\left(\frac{\left(f v_2^s\right)^2}{2}\left(2 \beta^2 +
 1 + \sqrt{8\beta^2 + 1 + \frac{8\beta}{f v_2^s}}\right) - 1\right)^{\frac{1}{2}}.
\end{equation}

This solution can now be plugged into \eqref{eq:h1_sol} and \eqref{eq:h2_sol2}
to find
\begin{subequations}
\begin{align}
 h_1 &= \frac{1}{4\beta}\left(\sqrt{8\beta^2 + 1 + \frac{8\beta}{f v_2^s}}  -
 1\right) \\ 
 h_2 &= \frac{1}{2} \frac{\Fc\kappa\left(3 +\sqrt{8\beta^2 + 1 +
 8\beta/f v_2^s}\right)} {f v_2^s\left(2\beta^2 + 1 + \sqrt{8\beta^2 + 1 + 8\beta/fv_2^s}\right) + 2 \beta}.
\end{align}
\end{subequations}
By definition, we had $h_3 = 1$. Finally, one has to replace $\beta$ by $v_1^s/v_2^s$ to get the result in terms of $v_1^s$ and $v_2^s$.

\section{Scaling law for the schematic model}
In this section, we give some arguments why the timescale for the decay close to the critical point scales like the inverse distance from the critical point. We start with the factorization ansatz $\phi_i(s) = h_i g(s)$ with the critical amplitudes $h_i$ and a general time dependency $g(s)$, which is inserted into the equations of motion close to the critical point given in Eq.~\eqref{eq:eom_schematic_critical}. One can identify the stability matrix $S^c$ (given in \eqref{eq:schematic:critical_stability_matrix}) so that the equations of motion read 
\begin{subequations}
\begin{align}
    g(s) \sum_i S^c(\vh)_{1i} h_i &= g(s) \Delta V_\parallel^\text{re}(\vh) + \tau_1 \\
    &\quad - s(g(s))^2(h_1^2-h_2^2) \notag\\
    g(s) \sum_i S^c(\vh)_{2i} h_i &= g(s) \Delta V_\parallel^\text{im}(\vh) + \tau_2\\
    &\quad - s(g(s))^2 2 h_1 h_2  \notag\\
    g(s) \sum_i S^c(\vh)_{3i} h_i &= 2 \Biggl(g(s)\left(\Delta \frac{\cW}{\cU}(\vh) \right) + \frac{\cV(\vh)}{\cU(\vh)} \\
    &\qquad - \tau_3 \frac{\cW(\vh)}{(\cU(\vh))^2} + \tau_3 - s(g(s))^2 h_3^2\Biggr)\notag
\end{align}
\end{subequations}
with the definitions 
\begin{subequations}
\begin{align}
    \Delta V_\parallel^\text{re/im}(\vh) &:= \sum_{i=1}^3 \left(\left(V_{\parallel}^\text{re/im}\right)_i - \left(V_{\parallel}^{\text{re/im},c}\right)_i\right) h_i, \\
    \cW(\vh) &:= \sum_{i,j=1}^3 (W_\perp)_{ij} h_i h_j, \\
    \cU(\vh) &:= \sum_{i=1}^3 (U_\perp)_i h_i, \\
    \cV(\vh) &:= \sum_{i=1}^3 (V_\perp)_i h_i, \\
   \Delta \frac{\cW}{\cU}(\vh) &:= \frac{\cW^c(\vh)}{\cU^c(\vh)} - \frac{\cW(\vh)}{\cU(\vh)},
\end{align}
\end{subequations}
where the superscript \emph{c} labels the vertices at the critical force. As the critical stability matrix $S^c$ is by definition not invertible at the critical point, this system of equations only has a solution if the right hand side lies in the invertible subspace (the image) of $S^c$. This can be achieved by adjusting the amplitude of $\vh$ such that the projection of the right hand side onto the left zero-eigenvector $\tilde{\vh}$ of $S^c$ vanishes. $\tilde{\vh}$ is given by the determining equation $\tilde{\vh}^T S^c = 0$ or equivalently $(S^c)^T \tilde{\vh} = 0$. Hence, the solution condition reads 
\begin{equation}
        0 = \tilde{c}_1 + g(s) \alpha \tilde{c}_2 + s(g(s))^2 \alpha^2 \tilde{c}_3
\end{equation}
with the abbreviations 
\begin{subequations}
 \begin{align}
     \tilde{c}_1 &= \tilde{h_1} \tau_1 + \tilde{h_2}\tau_2 + 2 \tilde{h_3} \left(\frac{\cV(\vh)}{\cU(\vh)} - \tau_3 \frac{\cW(\vh)}{(\cU(\vh))^2} + \tau_3\right),\\
     \tilde{c}_2 &= \tilde{h_1} \Delta V_\parallel^\text{re} (\vh) + \tilde{h_2} \Delta V_\parallel^\text{im}(\vh) + 2 \tilde{h}_3 \Delta \frac{\cW}{\cU}(\vh),\\
     \tilde{c}_3 &= \tilde{h_1}(h_1^2 - h_2^2) + \tilde{h_2} h_1 h_2 - 2 \tilde{h_3} h_3^2.
 \end{align}
\end{subequations}
This is a quadratic equation for $\alpha$, which can be solved explicitly. For the following, we assume that $\vh$ is normalized such that $\alpha=1$. Note that $\tilde{c}_3$ is independent of the external force and $\tilde{c}_2$ is proportional to the distance from the critical force to lowest order. Then, it remains to find a solution of the following equation (expanded to lowest order in $\delta$, the distance to the critical force)
\begin{equation}
\label{eq:schematic:betascalingequation}
    0 = c_1  + \delta c_2 g(s)  + s (g(s))^2 
\end{equation}
with $c_1 \tilde{c}_3 = \tilde{c}_1^c$
%, $c_1' = \partial_{\Fext} \tilde{c}_1 |_{\Fcrit} / \tilde{c}_3$ 
and $c_2 \tilde{c}_3 = \frac{\partial}{\partial {\Fext}} \tilde{c}_2|_{\Fc}$. 

This equation leads to the following scaling argument: 
%Given that $g_\pm(s)$ solves 
%\begin{equation}
%  \label{eq:schematic:scaling_equation}
%    0 = c_1 \pm c_2 g_\pm(s) + s (g_\pm(s))^2
%\end{equation}
%for every $s$. 
Inserting the ansatz $g(s):= \sigma g_{\pm}(\tau s)$ into \eqref{eq:schematic:betascalingequation} gives a scale-independent equation 
\begin{equation}
    0 = c_1 \pm c_2 \delta \sigma g_{\pm}(\tau s) + \frac{\sigma^2}{\tau} \tau s \left(g_{\pm}(\tau s)\right)^2
\end{equation}
%This equation reduces to Eq.~\eqref{eq:schematic:scaling_equation} 
if and only if $\sigma|\delta|=1$ and $\sigma^2 / \tau = 1$. $g_{\pm}(t)$ are then the solutions of this scale-independent equation. 
Hence, we conclude 
\begin{equation}
g(s)=|\delta|^{-1} g_{\pm}(s \delta^{-2}) \quad\text{ for }\delta \gtrless 0.
\label{eq:schematic:scaling_solution}
\end{equation}
This is the same scaling equation as in conventional schematic MCT models (see \cite{Schnyder2011,Franosch1997,Goetze2009}). 

\section{Scaling of the velocity memory kernel}
\label{sec:scaling_memory_kernel}
The schematic model exhibits a timescale $\tau = \delta^{-2}$ for the behavior close to the critical force. The same scaling can be observed in the numerical solution of the full model as shown in Fig.~\ref{rp-theo}. This justifies the assumption of a factorization of the correlators, i.e.\ $\vphi^s(t) = \vh g(s)$ with the same scaling function as for the schematic model, i.e. $g(s) = \delta^{-1} g_{+} (s \delta^{-2})$ (above the critical force, see Eq.~\eqref{eq:schematic:scaling_solution}) we find 
\begin{equation}
    \begin{split}
        &\int_0^\infty  m_0^{zz}(t') \du t' = 
        %\int_0^\infty m_0^{zz}[\vh g(t')] \du t' = m_0^{zz}[\vh] \int_0^\infty g(t')\du t' = m_0^{zz}[\vh] \int_0^\infty \frac{|\delta|}{\delta^2} g_+(t'/\delta^2) \du t'
        \lim_{s\to 0} m_0^{zz}(s) = \lim_{s\to 0} m_0^{zz}[\vh g(s)] \\
        =&\,  m_0^{zz}[\vh] \lim_{s\to 0} \delta^{-1} g_+(s \delta^{-2}) = m_0^{zz}[\vh] \delta^{-1} g_+(0),
    \end{split}
\end{equation}
where we used the linearity of the memory kernel functional $m_0^{zz}$ to separate the time dependency. This yields $\int_0^\infty m_0^{zz}(t')\du t' \propto |\delta|^{-1}$ close to the critical force.

\section{Stability matrix of the full model}
\label{chap:stability_matrix_full_model}
Separating the root problem
\eqref{eq:nonergodicity_equation_fp} into real and imaginary parts, we obtain 
\begin{subequations}
\begin{align}
J_q'^\re(\vv^\re,\vv^\im) &= v_q^\re - \frac{\are(\are+\bre)
+ \aim(\aim+\bim)}{(\are+\bre)^2+(\aim+\bim)^2}, \\
J_q'^\im(\vv^\re,\vv^\im) &= v_q^\im - \frac{\aim\bre -
\are\bim}{(\are+\bre)^2+(\aim+\bim)^2},
\end{align}
\end{subequations}
with $v_q^\re:=\Re v_q$, $v_q^\im=\Im v_q$ and 
\begin{subequations}
\begin{equation}
 \are = \Re \vv^T A_q \vv ,
\end{equation}
\begin{equation}
 \bre = \Re B_q^T \vv,
\end{equation}
\begin{equation}
 \aim = \Im \vv^T A_q \vv ,
\end{equation}
\begin{equation}
 \bim = \Im B_q^T \vv.
\end{equation}
\end{subequations}
The reduced stability matrix (i.e.\ the Jacobian of the non-trivial part) then consists of the four blocks 
\begin{widetext}
\begin{align}
 \pderiv{{J_q'}^\re}{v_p^{\re/\im}}  &= c\left(
 \left(\pderiv{\are}{v_p^{\re/\im}}\right)\left(
 		2\aim \bim \are - (\aim)^2\bre + \bre(\are + \bre)^2 + (\bim)^2(2\are+\bre)
 		\right) \right.\notag\\
 & + \left(\pderiv{\aim}{v_p^{\re/\im}}\right)\left(
 		\bim ((\aim)^2+(\bim)^2-(\are)^2+(\bre)^2) +
 		2\aim((\bim)^2+\bre(\are+\bre)) \right) \notag\\
 & + \left(\pderiv{\bre}{v_p^{\re/\im}}\right)\left(
 		-2\aim\bim\bre - (\aim)^2(\are+2\bre) - \are(-(\bim)^2+(\are+\bre)^2)
 		\right) \notag\\
 & \left.+ \left(\pderiv{\bim}{v_p^{\re/\im}}\right)\left(
 		-(\aim)^3 - 2(\aim)^2\bim - 2\bim \are(\are+\bre)
 		- \aim((\bim)^2+(\are)^2-(\bre)^2)
		 \right)
 \right)\\
 \pderiv{{J_q'}^\im}{v_p^{\re/\im}}  &= c\left(
 \left(\pderiv{\are}{v_p^{\re/\im}}\right)\left(
 		\bim((\are)^2-(\aim)^2 - (\bim)^2-(\bre)^2) -
 		2\aim((\bim)^2+\bre(\are+\bre)) \right)\right.\notag\\
 & + \left(\pderiv{\aim}{v_p^{\re/\im}}\right)\left(
 		2\aim \bim \are - (\aim)^2\bre + \bre(\are+\bre)^2
 		+ (\bim)^2 (2\are+\bre)
 		\right) \notag\\
 & + \left(\pderiv{\bre}{v_p^{\re/\im}}\right)\left(
 		(\aim)^3 + 2(\aim)^2\bim + 2\bim\are(\are+\bre)
 		+ \aim((\bim)^2+(\are)^2-(\bre)^2)
 		\right) \notag\\
 & \left.+ \left(\pderiv{\bim}{v_p^{\re/\im}}\right)\left(
 		-2\aim\bim\bre - (\aim)^2(\are+2\bre)
 		-\are(-(\bim)^2+(\are+\bre)^2)
 		\right)
 \right)
\end{align}
\end{widetext}
with $c=((\are)^2+(\aim)^2+(\bre)^2+(\bim)^2)^{-2}$. After calculating the
derivatives of $\are,\aim,\bre,\bim$ with respect to the variables $v_p^\re$ 
and $v_p^\im$, we are ready to construct the stability matrix as 
\begin{equation}
 S(\vv^\re,\vv^\im) = \mathbb{1} - \begin{pmatrix}
  \pderiv{{J_q'}^\re(\vv^\re,\vv^\im)}{v_p^\re}  
  & \pderiv{{J_q'}^\im(\vv^\re,\vv^\im)}{v_p^\re} \\
  \pderiv{{J'_q}^\re(\vv^\re,\vv^\im)}{v_p^\im}  
  & \pderiv{{J'_q}^\im(\vv^\re,\vv^\im)}{v_p^\im}
 \end{pmatrix}.
\end{equation}
Since $a$ and $b$ are holomorphic functions, there hold the Cauchy-Riemann differential equations as
long as $\vv\neq 0$
\begin{equation}
 \pderiv{{J_q'}^\re}{v_p^{\re}} = \pderiv{{J_q'}^\im}{v_p^{\im}} \qquad\text{and}\qquad
 \pderiv{{J_q'}^\re}{v_p^{\im}} = -\pderiv{{J_q'}^\im}{v_p^{\re}}.
\end{equation}

The largest eigenvalue of this stability matrix can be evaluated numerically and is real valued and approaches 0 at the delocalization transition. We also note that this eigenvalue has multiplicity 1. The
value of this largest eigenvalue coincides with the value of the contraction factor determined by the iteration. The eigenvector for this largest eigenvalue is also called the \emph{critical amplitude}
$\vh=(h_q)_q$ and it determines in classical mode-coupling theory the wave vector dependency of the
nonergodicity parameters close to the critical point via
\begin{equation}
\label{eq:beta:critical_wave_vector_scaling}
f_q-f_q^c=h_q g(\delta)
\end{equation}
(p.~239, Eq.~(4.78a) in \cite{Goetze2009}), where $g(\delta)$ describes the scaling behavior close to the critical force depending on the type of the transition: $g(\delta)\propto\sqrt{|\delta|}$ for a type B transition (p.~245, Eq.~(4.91a) in \cite{Goetze2009}) and $g(\delta)\propto \delta$ for a type A transition (p.~248, Eq.~(4.98) in \cite{Goetze2009}).  

In the case of vanishing nonergodicity parameters at the critical point, it is also possible to evaluate the eigenvector to the largest eigenvalue of the reduced stability matrix directly from the fixed point iteration scheme. Using \eqref{eq:beta:critical_wave_vector_scaling} with $f_q^c=0$ in the iteration equation \eqref{eq:nonergodicity_equation} and dividing by $g(\delta)$
we find in the limit $g(\delta)\to 0$ 
\begin{widetext}
\begin{equation}
\label{eq:beta:critical_amplitude_iteration}
 h_q = \frac{\are(\vh)\bre(\vh) - \aim(\vh)\bim(\vh) + \imath
 \left(\aim(\vh)\bre(\vh) - \are(\vh)\bim(\vh) \right)}{(\bre(\vh))^2 +
 (\bim(\vh))^2}.
\end{equation}
\end{widetext}

\end{document}